\RequirePackage[2020-02-02]{latexrelease}
\documentclass[preprint,aps,floatfix]{revtex4}
\DeclareUnicodeCharacter{00A0}{ }
\usepackage{graphicx}
\usepackage{textgreek}
\usepackage{verbatim} 
\usepackage{amssymb}
\usepackage{url}
\usepackage{amsmath,amssymb}
\usepackage{mathtools} 
\usepackage[symbol]{footmisc}
\usepackage{subcaption}
\usepackage{capt-of}
\usepackage{bigints}
\usepackage{romannum}
\usepackage{xcolor} 
\usepackage{soul} 
\AtBeginDocument{\pagenumbering{arabic}}

\def\be{\begin{eqnarray}}
\def\ee{\end{eqnarray}}
\def\nn{\nonumber}

\begin{document}

\title{Gluon Generalized TMDs and Wigner Distributions in boost invariant longitudinal space}

  \author{Sujit Jana${}^1$\footnote{d21ph010@phy.svnit.ac.in}, Vikash Kumar Ojha${}^1$\footnote{vko@phy.svnit.ac.in (corresponding author)}, and Tanmay Maji${}^2$\footnote{tanmayphy@nitkkr.ac.in}}
  \affiliation{${}^1$ Department of Physics, Sardar Vallabhbhai National Institute of Technology, Surat, 395 007, India.\\ ${}^2$ Department of Physics, National Institute of Technology, Kurukshetra, Haryana, 136 119, India.}
\begin{abstract}
We present the gluon generalized TMDs for non-zero skewness and Wigner distributions in the boost invariant longitudinal space. The boost-invariant longitudinal space is defined as $\sigma=\frac{1}{2}b^-\Delta^+$ and is conjugate to the skewness variable $\xi$. We use the dressed quark model, where a high-energetic quark is dressed by a gluon. This two-particle system has the advantage of addressing both the gluon and the quark sectors. The different contributions in Wigner distributions coming from different polarization of gluon and the dressed quark system are investigated. The Wigner distributions are obtained by taking the Fourier transformation of generalized TMDs, which we derive for the first time for non-zero skewness in dressed quark model.

\end{abstract}
\maketitle
\section{Introduction}
Proton and neutron are the bound state of quark and gluon held together by the strong force. The 3D distribution of quark and gluon in proton and their contribution to the physical properties of proton, like its mass, magnetic moment, and spin, are among the primary objectives
of the upcoming and exciting electron-ion collider (EIC) \cite{Accardi:2012qut}. The proposed electron-ion collider in China (EicC) also shares similar goals \cite{Anderle:2021wcy}.

Theoretically, the distribution of partons inside a nucleon with a very high longitudinal momentum is given by the 1-D parton distribution function (PDF) $f(x)$ \cite{Collins:1981uw,Martin:1998sq,Gluck:1994uf,Gluck:1998xa}. The PDF can provide only a 1-D view of the proton. To have a multidimensional view of the proton, we need multidimensional
function: generalized parton distribution (GPDs) $f(x,\xi,t)$ \cite{Ji:1996nm,Diehl:2003ny,Belitsky:2005qn,Goeke:2001tz} and transverse-momentum dependent PDFs (TMDs) $f(x,k^2_\perp)$ \cite{Mulders:1995dh,Barone:2001sp,Bacchetta:2006tn,Brodsky:2002cx,Bacchetta:2017gcc}. In the hierarchy of distribution functions, we have the generalized transverse momentum distribution function (GTMDs)\cite{Meissner:2008ay,Meissner:2009ww,Lorce:2013pza} at the top, along with its Fourier transform, which gives the Wigner distributions \cite{Wigner:1932eb,Ji:2003ak}. On the one hand, we have GTMDs, which reduces to GPDs, TMDs, and PDFs under specific limit. On the other hand, we have Wigner distribution, a quasi-probability distribution function, and a quantum mechanical analog of phase space in classical physics, and it becomes the probability density of partons under a certain limit\cite{Lorce:2011kd}. The Wigner distribution provides an excellent tool to investigate the system in both momentum and position space, not only in quantum chromodynamics but also in optics and quantum information\cite{Radhakrishnan:2022khp,PhysRevLett.109.190502,weinbub2018recent}.

The distribution functions GPDs, TMDs, GTMDs, and Wigner distributions has been studied extensively in recent times for the strongly bound system in the different model \cite{Mukherjee:2014nya,Mukherjee:2015aja,Lorce:2011kd,Lorce:2011ni,Kaur:2018dns,Liu:2015eqa,Maji:2022tog,Chakrabarti:2016yuw,Chakrabarti:2017teq,More:2017zqp,More:2017zqq,Mukherjee:2013yf}. Most studies in the last decade were focused on the quark distribution, but currently, the focus has shifted to the gluon sector, which is being studied intensely\cite{Hatta:2022bxn,PhysRevLett.129.252002,Bhattacharya:2023yvo,V:2023egw,Chakrabarti:2023djs,Lyubovitskij:2021qza,Boer:2022njw}. The GPDs can be extracted through the Deeply Virtual Compton Scattering (DVCS) and high energy meson production (HEMP) process\cite{Kouznetsov:2016vvo}, while the TMDs are experimentally accessible in the semi-inclusive DIS (SIDIS), Drell-Yan, and di-hadron production \cite{Boussarie:2023izj}. The GTMDs for quarks and gluons have been recently shown to be experimentally accessible through the exclusive double Drell-Yan process and in exclusive double production of pseudoscalar quarkonia, respectively \cite{Bhattacharya:2017bvs,Bhattacharya:2018lgm}.

Most of the previous analyses are done by considering the momentum transferred to the target state to be purely in the transverse direction \cite{Mukherjee:2014nya,Mukherjee:2015aja,Zhou:2016rnt,Kaur:2018dns,Chakrabarti:2017teq,Chakrabarti:2016yuw}. In this article, we explore the gluon GTMDs by 
considering the impact of a non-zero momentum transferred to the target state in the longitudinal direction. The variable $\xi=(p-p^{'})/(p+p^{'})$, commonly known as skewness, encapsulates the longitudinal-momentum transferred to the target state. Taking the Fourier transform of GTMDs gives the Wigner distribution in the boost-invariant longitudinal ($\sigma$) space. The boost-invariant longitudinal space ($\sigma$ space) is defined as
the spatial coordinate conjugate to the longitudinal momentum transferred to the target state \cite{Brodsky:2006in}. The articles \cite{Brodsky:2006ku,Chakrabarti:2005zm,Chakrabarti:2015ama,Manohar:2010zm,Mukherjee:2011an} explored the GPDs in σ-space and the impact parameter space to get a 3D view of the hadron, while \cite{Maji:2022tog,Ojha:2022fls} studied the Wigner distribution for quark in σ-space. This article investigates the Wigner distribution for gluon in the boost invariant longitudinal space.

\section{Convention and Kinematics \label{sec:Conven}}
We opt for a light-front coordinate system denoted as $(x^+, x^-, x_\perp)$, where the light-front time variable ($x^+$) and the light-front longitudinal space variable ($x^-$) are defined as $x^{\pm} = x^0 \pm x^3 $ \cite{Zhang:1994ti,Harindranath:1996hq}. Our focus involves a system comprising a quark dressed with a gluon, subjected to probing by a virtual photon. The total squared momentum transferred from the virtual photon to the target state is denoted as $t=(p-p^\prime)^2=\Delta^2$. The longitudinal momentum transferred to the target is expressed through the skewness variable $\xi$, defined as $\xi = \Delta^+/2P^+$. We adopt a symmetric frame and parameterize the initial and final four-momentum of the target state according to the reference \cite{Brodsky:2000xy} as
\be
    p &=&\Bigg((1+\xi)P^+,\Delta_\perp/2,\frac{m^2+\Delta_\perp^2/4}{(1+\xi)P^+}\Bigg) ;\\
    p'&=&\Bigg((1-\xi)P^+,-\Delta_\perp/2,\frac{m^2+\Delta_\perp^2/4}{(1-\xi)P^+}\Bigg),
\ee
where $P=(p + p^\prime)/2$ is the average momentum of the target state and the  momentum transferred to the target state is 
\be
    \Delta = p-p' = \Bigg(2\xi P^+, \Delta_\perp, \frac{t+\Delta_\perp^2}{2\xi P^+} \Bigg),
\ee
with
\be
    t=-\frac{4\xi^2 m^2 + \Delta_\perp^2}{1-\xi^2}.
    \ee
The gluon carries the average longitudinal momentum fraction $x_g=k^+_g/P^+$ of the target state and the gluon four-momentum is
\be 
k_g\equiv \Big(x_g P^+, k_{\perp g}, k^-_g \Big)\label{mom_k}.
\ee

\section{Light-front Dressed Quark Model}
To study the GTMDs, we are considering a dressed quark as the target state. A dressed quark can be considered as a bound-state system with spin $1/2$, consisting of a bare quark and a gluon. The model has been used previously to study the behavior of quark and gluon in a bound-state system \cite{Harindranath:1998pc,Harindranath:1998ve,Ojha:2022fls,Mukherjee:2015aja,Mukherjee:2014nya,mukherjee2015wigner,More:2023pcy}. A dressed quark state with momentum $p$ and spin $\sigma$ can be expanded in the Fock space as \cite{Harindranath:1998pc,Harindranath:1998ve}, 
\be \label{fockse}
  \Big{| }p^{+},p_{\perp},\sigma  \Big{\rangle} = \Phi^{\sigma}(p) b^{\dagger}_{\sigma}(p)
 | 0 \rangle +
 \sum_{\sigma_1 \sigma_2} \int [dp_1]
 \int [dp_2] \sqrt{16 \pi^3 p^{+}}
 \delta^3(p-p_1-p_2) \nn \\ \Phi^{\sigma}_{\sigma_1 \sigma_2}(p;p_1,p_2) 
b^{\dagger}_{\sigma_1}(p_1) 
 a^{\dagger}_{\sigma_2}(p_2)  | 0 \rangle;
\ee
where, we truncated the series expansion up to two-particle state. Here, $[dp] =  \frac{dp^{+}d^{2}p_{\perp}}{ \sqrt{16 \pi^3 p^{+}}}$, $\Phi^\sigma(p)$ is the single-particle wavefuction with momentum $p$ and spin $\sigma$. $\Phi^{\sigma}_{\sigma_1 \sigma_2}(p;p_1,p_2)$ is the two-particle light front wave function and can be expressed in the boost-invariant form using the relation  $\Psi^{\sigma}_{\sigma_1
\sigma_2}(x, q_\perp) =   
\Phi^{\sigma}_{\sigma_1 \sigma_2}
\sqrt{P^+}$.
The momentum variables $(x_i, q_{i \perp})$ are Jacobi momenta defined as 
\be \label{jacmom}
p_i^+= x_i p^+, ~~~~~~~~~~q_{i \perp}= k_{i \perp}+x_i p_\perp,
\ee
and satisfies the following constraints 
\begin{equation*}
    \sum_i x_i=1, ~~~~~~~~~\sum_i q_{i\perp}=0
\end{equation*}
Taking $(x_1,q_{1\perp})\equiv (x,q_\perp)$ for quark, and $(x_2,q_{2\perp})\equiv(x_g,q_{\perp g})$ for gluon, we get
 \begin{align}
x+x_g=&1\Rightarrow x_g=1-x,\\
q_{\perp}+q_{\perp g}=&0 \Rightarrow q_{\perp g}=- q_{\perp}.
\end{align}
Therefore, the boost invariant two-particle LFWFs defined using momentum of gluon read \cite{PhysRevD.59.116013}

\be \label{tpag}
\Psi^{\sigma a}_{\sigma_1 \sigma_2}(x_g,q_{\perp g}) = 
\frac{1}{\Big[    m^2 - \frac{m^2 + (q_{\perp g})^2 }{x_g} - \frac{(q_{\perp g})^2}{1-x_g} \Big]}
\frac{g}{\sqrt{2(2\pi)^3}} T^a \chi^{\dagger}_{\sigma_1} \frac{1}{\sqrt{1-x_g}}
\nn \\ \Big[ 
-2\frac{q_{\perp g}}{1-x_g}   -  \frac{(\sigma_{\perp}.q_{\perp g})\sigma_{\perp}}{x_g}
+\frac{im\sigma_{\perp}(1-x_g)}{x_g}\Big]
\chi_\sigma (\epsilon_{\perp \sigma_2})^{*}.
\ee
Here, $T^a$ is the $SU(3)$ color matrices, m is the mass of dressed quark, $\chi$ is the 2-component spinor, and $\epsilon_\perp$ is the polarization vector of gluon. We choose initial and final Jocbi momenta for the target state as $(y,q_\perp)$ and $(x',q'_\perp)$, respectively. Using the above wave functions, this model has the liberty to investigate the non-perturbative distributions for gluon e.g, Generalised parton distribution (GPDs), transverse momentum distributions (TMDs), Generalised transverse momentum (GTMDs) etc. In this work, we concentrate on the leading twist GTMDs for gluons and some of the Wigner distributions.  
\section{Gluon Generalized TMDs}
GTMDs for gluon are the most general correlation function, encode the maximum amount of information, and are defined via the parametrization of off-forward gluon-gluon correlator. 
The general gluon-gluon correlator function $W^{[\Gamma]}_{\lambda,\lambda^\prime} (x,\xi,{k_{\perp}},{\Delta_{\perp}}) $ is defined as \cite{Lorce:2013pza}
\be \label{qqc}
W^{\mu\nu;\rho\sigma}_{\lambda,\lambda^\prime} (P,x_g,\vec k_\perp,\Delta,N;\eta,\eta')
=\frac{1}{x_gP^+}\int\frac{dz^{-}}{2\pi}\frac{d^{2} z_{\perp}}{(2\pi)^2}e^{ix_gP^+z^--i\vec k_\perp\cdot\vec z_\perp}
 \Big{\langle } p^{\prime},\lambda^{\prime} \Big{|}
2Tr\Big[G^{\mu\nu}(-z/2) \mathcal{W}\nn \\
G^{\rho\sigma}(z/2)\mathcal{W}^{\prime}\Big] \Big{|}
p,\lambda \Big{\rangle }
\Big{|}_{z^{+}=0},
\ee 
where, the initial ${|}p,\lambda {\rangle } $ and final state ${|} p^{\prime},\lambda^{\prime}\rangle $ of the dressed quark system can be expanded in the Fock space using Eq.(\ref{fockse}) to express the gluon correlator in terms of the light-front wave function. $G^{\mu \nu}$ and $G^{\rho \sigma}$ are the gluon field strength tensor at two different spatial points having a connection through the Wilson lines $\mathcal{W}$ and $\mathcal{W^\prime}$ which preserve the Lorentz invariance. 
We choose the light-front gauge $A^ + = 0$ in which the Wilson line $\mathcal{W}$ and $\mathcal{W}^\prime$ reduces to $1$, and the gluon-gluon correlator reads
\be \label{eq:qqc LF Gauge}
W^{[\Gamma^{ij}]}_{\lambda,\lambda^\prime} (P,x_g,\vec k_\perp,\Delta,N;\eta,\eta')
=\frac{1}{x_gP^+}\int\frac{dz^{-}}{2\pi}\frac{d^{2} z_{\perp}}{(2\pi)^2}e^{ix_gP^+z^--i\vec k_\perp\cdot\vec z_\perp}\nn \\
 \Big{\langle } p^{\prime},\lambda^{\prime} \Big{|}
\Gamma^{ij} G^{+i}\Big( -\frac{z}{2}\Big) G^{+j}\Big( \frac{z}{2}\Big)
 \Big{|}
p,\lambda \Big{\rangle }
\Big{|}_{z^{+}=0}
\ee 
with the gluon field strength tensor
\be
G^{+i} = \partial^+ A^i -  \partial^i  A^+ + g f_{abc} A^{+}A^{i}
 \ee
reducing to $G^{+i} = \partial^+ A^i$. The different polarization contributions of gluon can be projected out using $\Gamma^{ij}=\{ \delta^{ij}_\perp, -i\epsilon^{ij}_\perp \}$. 
In this article we focus on the unpolarized ($\Gamma=\delta^{ij}_\perp$) and longitudinally polarized ($\Gamma=-i\epsilon^{ij}_\perp$) gluon. In Eq.(\ref{eq:qqc LF Gauge}), $x_g$ is the momentum fraction carried by the gluon in the dressed quark system, whereas $k_\perp(=-k_{\perp g})$ is the transverse momentum of the quark and $P^+$ is the average longitudinal momentum of the target state defined as $P^+=\frac{p^+ +p^{\prime +}}{2}=\frac{k^+}{x_g}$.
\noindent
The transverse component of gauge field for $i=1,2$ can be written as \cite{PhysRevD.59.116013}
\be
A^{i}\Big(\frac{z}{2}\Big) = 
\sum _{\lambda} \int \frac{dk^{+}d^{2} k_{\perp}}{2k^{+}(2\pi)^3}  
\Big[  \epsilon_{\lambda}^{i}(k) a_{\lambda}(k)e^{-\frac{i}{2}k.z} +   
\epsilon_{\lambda}^{*i}(k) a^{\dagger}_{\lambda}(k)e^{\frac{i}{2}k.z} \Big]. \ee
\noindent
%
%
%
 Truncating the Fock space expansion up to two particle state, we get the gluon-gluon correlator for unpolarized and longitudinally polarized gluons as
\be
     W^{(\delta^{ij}_\perp)}_{\sigma\sigma'}(x_g,\xi,{ k_{\perp}},{\Delta_{\perp}}) &= -\displaystyle\sum_{{\sigma_{1}},{\lambda_{1}},{\lambda_2}}[\Psi_{\sigma_{1}\lambda_{1}}^{{*}{\sigma'}}{(x'_g,q'_{\perp g})}\;{{\Psi_{{\sigma_{1}}{\lambda_{2}}}^{{\sigma}}}{(y_g,q_{\perp g})}}(\epsilon^1_{\lambda_2}\epsilon^{*1}_{\lambda_1}+\epsilon^2_{\lambda_2}\epsilon^{*2}_{\lambda_1})] \label{ggc-unpolarized}\\
      W^{(-i\epsilon^{ij}_\perp)}_{\sigma\sigma'}(x_g,\xi,{ k_{\perp}},{\Delta_{\perp}}) &= 
   -i\displaystyle\sum_{{\sigma_{1}},{\lambda_{1}},{\lambda_2}}[\Psi_{\sigma_{1}\lambda_{1}}^{{*}{\sigma'}}{(x'_g,q'_{\perp g})}\;{{\Psi_{{\sigma_{1}}{\lambda_{2}}}^{{\sigma}}}{(y_g,q_{\perp g})}}(\epsilon^1_{\lambda_2}\epsilon^{*1}_{\lambda_1}-\epsilon^2_{\lambda_2}\epsilon^{*2}_{\lambda_1})],\label{ggc-long-polarized}
\ee
where $(y_g, q_{\perp g}),\;(x_g^\prime,q_{\perp g}^\prime)$ are the initial and final momentum of the gluon, parameterized as \cite{Brodsky:2000xy} 
\be
    x'_g&=1-\frac{x-\xi}{1-\xi}=\frac{x_g}{1-\xi} ~~,~~~
    q'_{\perp g}=-k_\perp-\frac{(1-x)}{(1-\xi)}\frac{\Delta_\perp}{2}=-k_\perp-\frac{x_g\Delta_\perp}{2(1-\xi)}\nn\\
    y_g&=1-\frac{x+\xi}{1+\xi}=\frac{x_g}{1+\xi} ~~,~~~
    q_{\perp g}=-k_\perp+\frac{(1-x)}{(1+\xi)}\frac{\Delta_\perp}{2}=-k_\perp+\frac{x_g\Delta_\perp}{2(1+\xi)}\nn
\ee


There are two different but equivalent bilinear decomposition of quark-quark correlators are possible in terms of GTMDs \cite{Meissner:2009ww,Lorce:2013pza}. However, the decomposition of \cite{Lorce:2013pza} is also valid for the gluon correlator and we use it to obtain the analytical expression of gluon GTMDs. The gluon-gluon correlator Eq.(\ref{eq:qqc LF Gauge}) can be written compactly as \cite{Lorce:2013pza}
\begin{align}
    W_{\lambda,\lambda'}^{\Delta S_z,c_p}=\frac{\Bar{u}(p',\lambda')M^{\Delta S_z,c_p}u(p,\lambda)}{2P^+\sqrt{1-\xi^2}},
\end{align}
where $M^{\Delta S_z,c_p}$ is the matrix in the Dirac space corresponding the the gluon operator, $c_p$ is the parity coefficient of gluon operator and $\Delta S_z$ is the spin flip number defined as $\Delta S_z=\lambda'-\lambda+\Delta L_z$. Since $\Delta L_z$, the eigenvalue of operator $\Delta {\hat{L_z}}=\hat{L_z}-\hat{L'_z}$ is zero for twist-2 operators, the spin-flip number for twist-2 operators is just the difference of helicity of final and initial states. The exact parametrization of gluon-gluon correlator for the gluon operator $\delta^{ij}G^{+i}(-z/2)G^{+j}(z/2)$ and $-i\epsilon_\perp^{ij}G^{+i}(-z/2)G^{+j}(z/2)$ are \cite{Lorce:2013pza}
\begin{align}
    {W^{[\delta^{ij}]}_{\lambda,\lambda'}}&=\frac{1}{2P^+ \sqrt{1-\xi^2}}\Bar{u}(p',\lambda')\Big[\gamma^+(S^{0,+}_{1,1a}+\gamma_5\frac{i\epsilon^{k_T\Delta_T}_T}{m^2}S^{0,+}_{1,1b})+i\sigma^{j+}(\frac{k^j_T}{m}P^{0,+}_{1,1a}+\frac{\Delta^j_T}{m}P^{0,+}_{1,1b})\Big]u(p,\lambda),\label{bilinear decomposition unpol}\\
      W^{[-i\epsilon^{ij}_\perp]}_{\lambda,\lambda'}&=\frac{1}{2P^+ \sqrt{1-\xi^2}}\Bar{u}(p',\lambda')\Big[\gamma^+\gamma_5(S^{0,-}_{1,1a}+\gamma_5\frac{i\epsilon^{k_T\Delta_T}_T}{m^2}S^{0,-}_{1,1b})+i\sigma^{j+}\gamma_5(\frac{k^j_T}{m}P^{0,-}_{1,1a}+\frac{\Delta^j_T}{m}P^{0,-}_{1,1b})\Big]u(p,\lambda)\label{bilinear decomposition Long}
  \end{align}
The GTMDs $X=\{S^{0,+}_{1,1a}, S^{0,+}_{1,1b},P^{0,+}_{1,1a},P^{0,+}_{1,1b},S^{0,-}_{1,1a},S^{0,-}_{1,1b},P^{0,-}_{1,1a},P^{0,-}_{1,1b}\}$ are the function of
$(x_g,\xi,{ k^2_\perp},{\Delta^2_\perp},{k_\perp\cdot\Delta_\perp})$ and satisfies the hermiticity constraint
\be
X^*(x_g,\xi,{ k^2_\perp},{\Delta^2_\perp},{k_\perp\cdot\Delta_\perp})=\pm X(x_g,-\xi,{ k^2_\perp},{\Delta^2_\perp},{-k_\perp\cdot\Delta_\perp})\label{hermiticity condition}.
\ee
Within the dressed quark model, the gluon-gluon correlators provided in Eq. (\ref{ggc-unpolarized},\ref{ggc-long-polarized}) can be computed by utilizing the two-particle light-front wavefunction (Eq.(\ref{tpag})). By substituting the evaluated expression for gluon-gluon correlators into Eq. (\ref{bilinear decomposition unpol},\ref{bilinear decomposition Long}), we derive the analytical expression for GTMDs as follows.
(a) For unpolarized gluon, 
    \begin{align}
  S^{0,+}_{1,1a}&=-\frac{\alpha_g}{2x_g}\Big[4m^2((1-x_g)^2-\xi^2)^2+(1+x_g^2-\xi^2)(4(1-\xi^2)k^2_\perp-x_g^2\Delta^2_\perp+4x_g\xi k_\perp\cdot\Delta_\perp)\Big]\label{gtmd-unp-1}\\
           S^{0,+}_{1,1b}&=\alpha_g\Big[2m^2(x_g^2+\xi^2-1)\Big]\label{gtmd-unp-2}\\
           P^{0,+}_{1,1a}&=\beta_g\Big[2m^2(2\xi(x_g^2+\xi^2-1)k_\perp\cdot\Delta_\perp-x_g((1-x_g)^2+\xi^2)\Delta^2_\perp)\Big] \label{gtmd-unp-3}\\
           P^{0,+}_{1,1b}&=\beta_g\Big[-2m^2(2\xi(x_g^2+\xi^2-1)k^2_\perp-x_g((1-x_g)^2+\xi^2)k_\perp\cdot\Delta_\perp)\Big] \label{gtmd-unp-4}
    \end{align}
 (b) For Longitudinally Polarized gluon,
 \begin{align}
          S^{0,-}_{1,1a}&=-\frac{\alpha_g}{2x_g}\Big[4m^2((1-x_g)^2-\xi^2)^2+(1-x_g^2-\xi^2)(4(1-\xi^2)k^2_\perp-x_g^2\Delta^2_\perp+4x_g\xi k_\perp\cdot\Delta_\perp)\Big]\label{gtmd-lnp-1}\\
           S^{0,-}_{1,1b}&=-\alpha_g\Big[2m^2(x_g^2-\xi^2+1)\Big]\label{gtmd-lnp-2}\\
           P^{0,-}_{1,1a}&=-\beta_g\Big[4m^2(((1-x_g)^2-\xi^2(1-2x_g))k_\perp\cdot\Delta_\perp+x_g\xi(1-x_g)\Delta^2_\perp)\Big] \label{gtmd-lnp-3}\\
           P^{0,-}_{1,1b}&=\beta_g\Big[4m^2(((1-x_g)^2-\xi^2(1-2x_g))k^2_\perp+x_g\xi(1-x_g)k_\perp\cdot\Delta_\perp)\Big]\label{gtmd-lnp-4}
  \end{align}
 where we define the function $D(k_\perp,x_g)$, and $\beta_g(x_g,\xi, k^2_\perp,\Delta^2_\perp, k_\perp\cdot\Delta_\perp)$ as 
\begin{align}
    D(k_\perp,x_g)=&\Bigg(m^2-\frac{m^2+k_\perp^2}{x_g}-\frac{k^2_\perp}{1-x_g}\Bigg)~,~~~ \beta_g \equiv \frac{\alpha_g}{(k_2\Delta_1-k_1\Delta_2)}, \nn\\
   \text{where,~~} &\alpha_g(x_g,\xi, k^2_\perp,\Delta^2_\perp, k_\perp\cdot\Delta_\perp)=\frac{N\sqrt{1-\xi^2}}{D(q_{\perp g},y_g)D^*(q'_{\perp g},x'_g)x_g((1-x_g)^2-\xi^2)^\frac{3}{2}},
    \end{align}
and $N=\frac{g^2 C_f}{2(2\pi)^3}$. Here $C_f$ is the color factor and $g$ is the strong coupling constant. 
Note that the analytical expression of all GTMDs for gluon are real-valued functions and satisfy the hermiticity condition (Eq.(\ref{hermiticity condition})). That is, under the symmetry transformation $\xi\rightarrow -\xi$, and $k_\perp\rightarrow -k_\perp$, 
\be
X(x_g,\xi,{ k^2_\perp},{\Delta^2_\perp},{k_\perp\cdot\Delta_\perp})=+ X(x_g,-\xi,{ k^2_\perp},{\Delta^2_\perp},{-k_\perp\cdot\Delta_\perp})
\ee
 for the GTMDs $X=\{S^{0,+}_{1,1a},S^{0,+}_{1,1b},P^{0,+}_{1,1b}, S^{0,-}_{1,1a},S^{0,-}_{1,1b},P^{0,-}_{1,1a}\}$, and
\be
X(x_g,\xi,{ k^2_\perp},{\Delta^2_\perp},{k_\perp\cdot\Delta_\perp})=- X(x_g,-\xi,{ k^2_\perp},{\Delta^2_\perp},{-k_\perp\cdot\Delta_\perp})
\ee
 for the  GTMDs  $X=\{P^{0,+}_{1,1a},P^{0,-}_{1,1b}\}$.
In general, the GTMDs are imaginary, and can be split into two parts: odd ($X^o$) and even ($X^e$),
\be
X=X^e+i X^o,
\ee
 with both odd and even part being the real-valued functions. 

In the case of quark GTMDs, model calculations indicate that, up to the lowest order, all T-odd components of the GTMDs vanish, resulting in the GTMDs being real-valued functions \cite{Meissner:2009ww,Mukherjee:2014nya,Ojha:2022fls}. Similarly, we observe that for the lowest order calculation, gluon GTMDs (Eq.(\ref{gtmd-unp-1}-\ref{gtmd-lnp-4})) are also real-valued functions, mirroring the behavior of quark GTMDs. This suggests that the T-odd part of gluon GTMDs also vanishes in the lowest order calculation within our model. Appendix-B provides the analytical expression of gluon GTMDs in the notation of \cite{Meissner:2009ww}.
The analytical results in this model (Eqs.(\ref{gtmd-unp-1}-\ref{gtmd-lnp-4}), Eqs.(\ref{eq: gluon F11}-\ref{eq: gluon G14})) are used to investigate several properties of gluon and are presented through numerical plots.
\subsection{Orbital angular momentum of gluon}
The determination of spin distribution in strongly bound states such as protons or neutrons remains an unresolved issue and stands as a primary objective for the future Electron-Ion Collider (EIC) \cite{Aschenauer:2015ata,deFlorian:2019egz,Ji:2020ena}. Theoretical investigations into proton spin decomposition are characterized by two major approaches: the kinetic decomposition proposed by Ji and the canonical decomposition introduced by Jaffe and Manohar. The significance of GTMDs is also underscored by their direct relevance to both the kinetic and canonical definitions of the orbital angular momentum of partons. Expressions for spin ($S^g$), orbital angular momentum ($l^{g}_{z}$), and spin-orbit correlation ($C^{g}_{z}$) for gluons in terms of GTMDs can be found in the literature as \cite{Lorce:2011ni,Ji:1996nm,Leader:2013jra} 
\begin{align}
 S^g=\int dx d^2k_\perp G^g_{1,4}(x_g,0,{ k^2_\perp},0,0), \label{spinG} \\ 
 l^{g}_{z} = -\int dx d^{2}k_{\perp} \frac{k_{\perp}^2}{m^2} F^g_{14}(x_g,0,{ k^2_\perp},0,0),\label{oamG}\\
 C^{g}_{z} = \int dx d^{2}k_{\perp} \frac{k_{\perp}^2}{m^2} G^g_{11}(x_g,0,{ k^2_\perp},0,0)\label{socG}.
\end{align}
The integration range for $k_\perp$ ideally extends from $0$ to $\infty$. However, in accordance with standard numerical integration procedures, we adopt an upper cutoff $Q$ and a lower cutoff $\mu$. 

The lower cutoff $\mu$ is set to $0$, and for the upper cutoff $Q=50$ GeV, we obtain $S^g=-0.177$ GeV, $l^{g}_{z}=-0.020$ GeV, and $C^{g}_{z}=0.302$ GeV. The variable $x$ is integrated over the range 0 to 0.99 in Eq.(\ref{spinG},\ref{oamG}) and 0.01 to 0.99 in Eq.(\ref{socG}) to avoid singularities at $x=0$ and $x=1$. In our model, where $C^{g}_{z}>0$, the gluon orbital angular momentum (OAM) and spin are aligned. However, the gluon OAM is anti-aligned with the spin of the dressed quark as $l^{g}_{z}<0$.

\subsection{TMD Limit}
The GTMDs may not have direct physical implications, but their limit to TMDs or other lower distribution provides physical information about the internal structure of the target.
GTMDs map directly to the TMDs in the forward limit $\xi=0,~ \Delta^2_\perp=0$ and $k_\perp\cdot\Delta_\perp=0$ \cite{Meissner:2009ww,Lorce:2013pza}. The analytical results for the TMDs $f^g_1$, $f_{1T}^{\perp g}$, $g_{1L}^g$, and $g_{1T}^g$ of gluon read as   
\begin{align}
f^g_1(x_g,\Vec{k}_\perp^2)&=\mathrm{Re}[S^{0,+}_{1,1a}(x_g,0,\Vec{k}_\perp^2,0,0)]=\frac{2N(k_\perp^2(1+x_g^2)+m^2(1-x_g)^4)}{(1-x_g)(k_\perp^2+m^2(1-x_g)^2)^2}   \\
f_{1T}^{\perp g}(x_g,\Vec{k}_\perp^2)&=-\mathrm{Im}[P^{0,+}_{1,1a}(x_g,0,\Vec{k}_\perp^2,0,0)]=0\\
g_{1L}^g(x_g,\Vec{k}_\perp^2)&=\mathrm{Re}[S^{0,-}_{1,1a}(x_g,0,\Vec{k}_\perp^2,0,0)]=-\frac{2N(m^2(1-x_g)^3+k_\perp^2(1+x_g))}{(k_\perp^2+m^2(1-x_g)^2)^2}\\
g_{1T}^g(x_g,\Vec{k}_\perp^2)&=\mathrm{Re}[P^{0,-}_{1,1a}(x_g,0,\Vec{k}_\perp^2,0,0)]=0.
\end{align}

\begin{figure}[!htp]
\begin{minipage}[c]{1\textwidth}
\small{(a)}\includegraphics[width=7cm,height=4.9cm,clip]{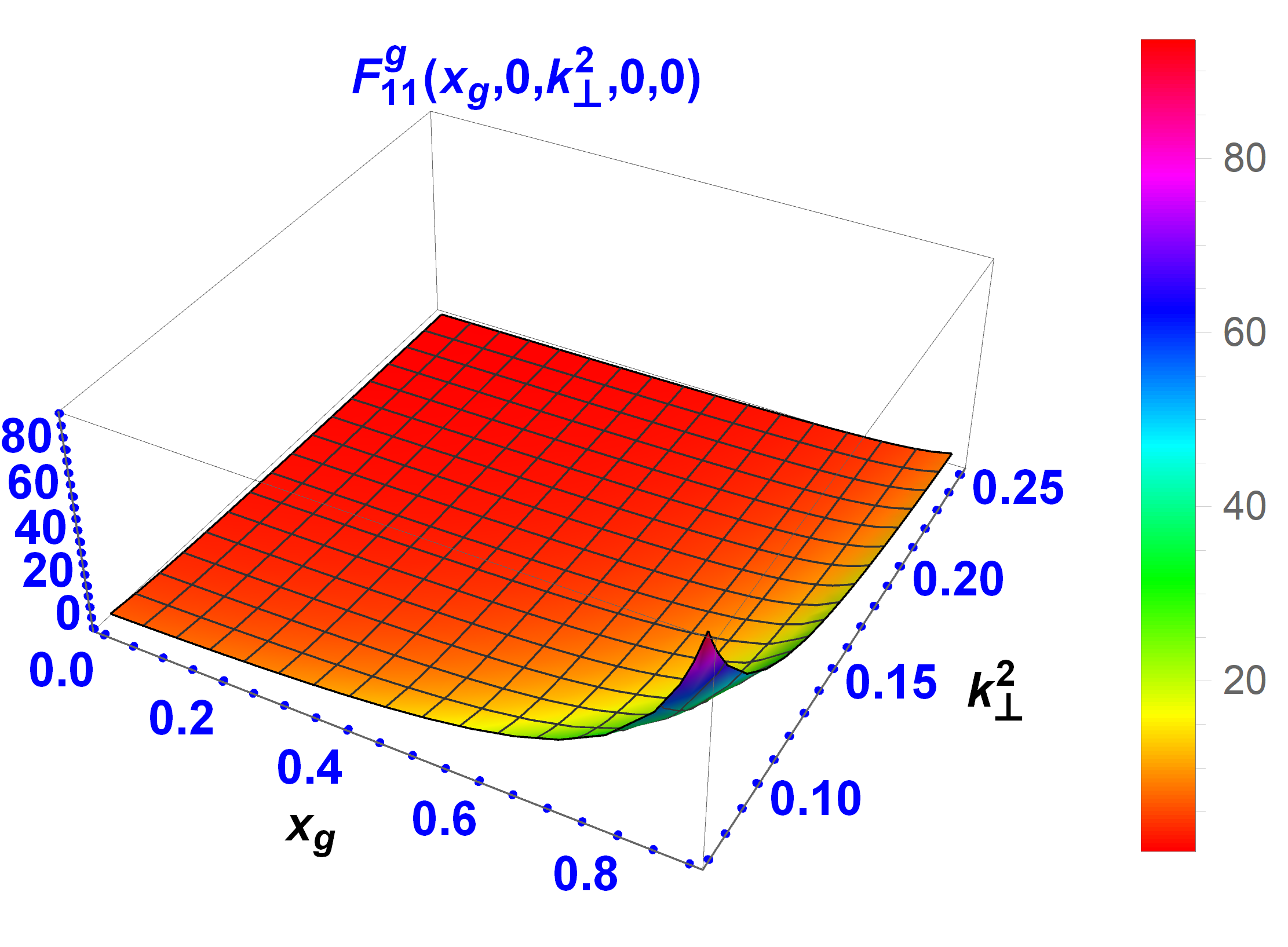}
\hspace{0.1cm}
\small{(b)}\includegraphics[width=7cm,height=4.9cm,clip]{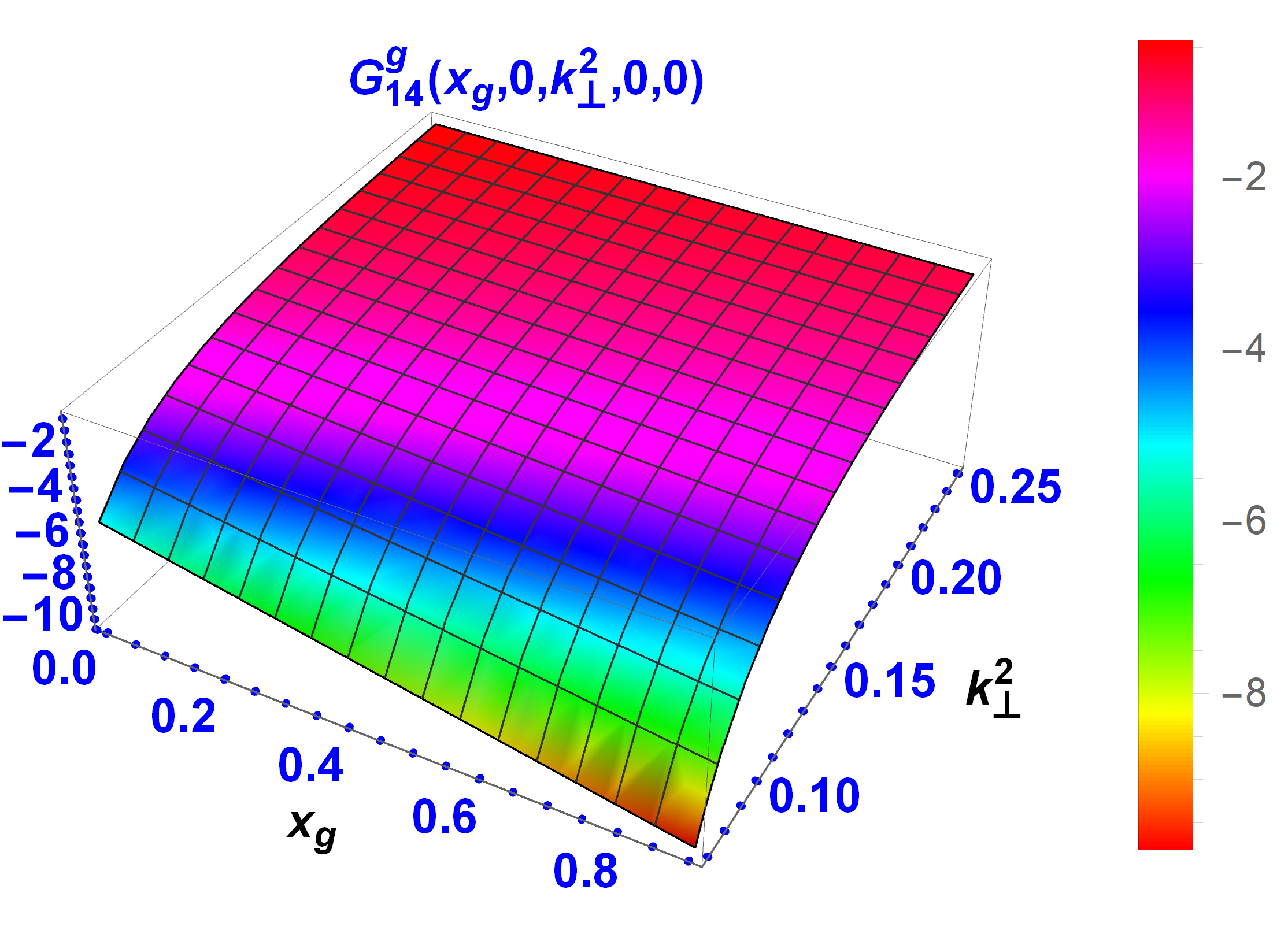} \\
\end{minipage}
\caption{  \label{fig_TMD} Plot of GTMDs (a) $F^g_{11}$ and (b) $G^g_{14}$ in the forward limit $\xi=0,~ \Delta^2_\perp=0,~k_\perp\cdot\Delta_\perp=0$, and $m=3.3$ MeV.}
\end{figure}

\begin{figure}[!htp]
\begin{minipage}[c]{1\textwidth}
\small{(a)}\includegraphics[width=7cm,height=4.9cm,clip]{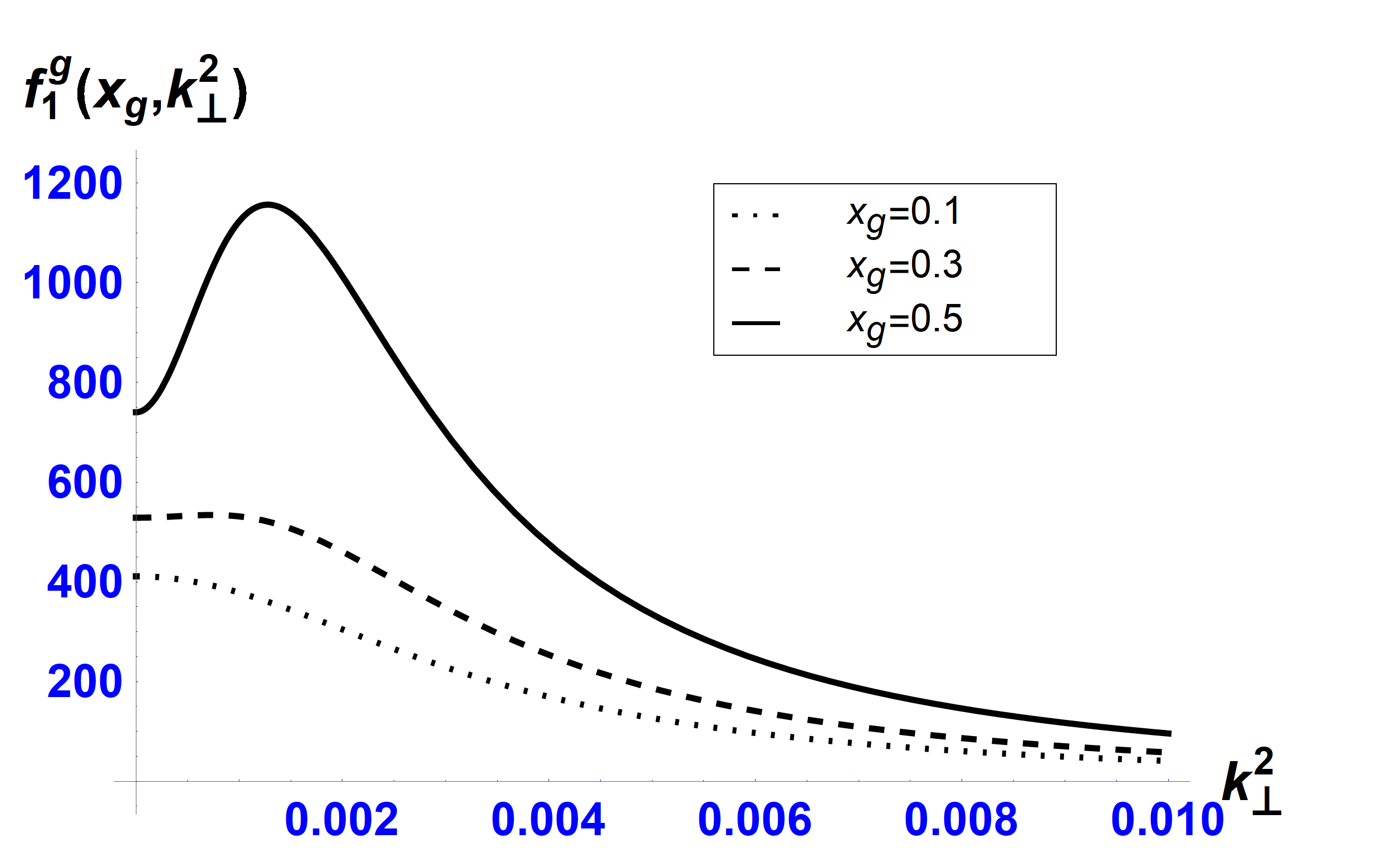}
\hspace{0.1cm}
\small{(b)}\includegraphics[width=7cm,height=4.9cm,clip]{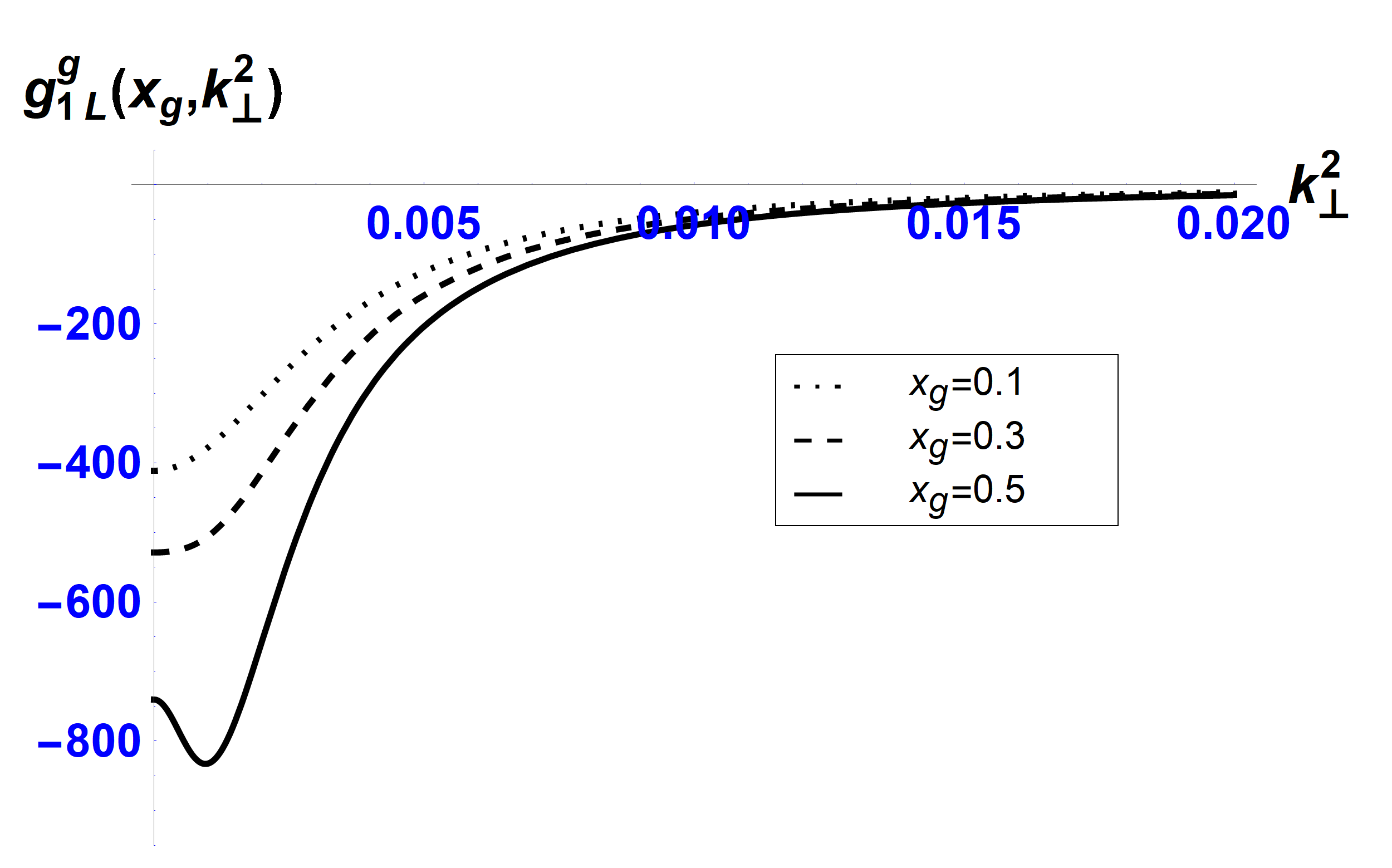} \\
\end{minipage}
\caption{\label{fig_TMD_kp} Plot of TMD (a)$f^g_{1}(x,k^2_\perp)$ and (b) $g^g_{1L}(x,k^2_\perp)$ as the function of $k_\perp^2$ for $x=0.1, 0.3, 0.5$, and $m=3.3$ MeV.}
\end{figure}

TMD $f_{1T}^{\perp g}(x_g,\Vec{k}_\perp^2)$ and $g_{1T}^g(x_g,\Vec{k}_\perp^2)$ vanishes in our model at the leading order. However, the higher order calculation may results in a non-zero expression for these two TMDs. The non vanishing TMDs $f^g_1(x_g,\Vec{k}_\perp^2)$ and $g_{1L}^g(x_g,\Vec{k}_\perp^2)$ are presented in the three-dimension plot with $x_g$ and $\Vec{k}_\perp^2$ in Fig.\ref{fig_TMD}. The $\Vec{k}_\perp^2$ variation of the GTMDs are also shown in the low $\Vec{k}_\perp^2$ region in Fig.\ref{fig_TMD_kp}.


\subsection{Helicity density and Worm-gear density}
\begin{figure}%
    \centering
    \subfloat[\centering $x\rho^{\circlearrowleft/+}_g(x,k_\perp)$]{{\includegraphics[scale=0.68]{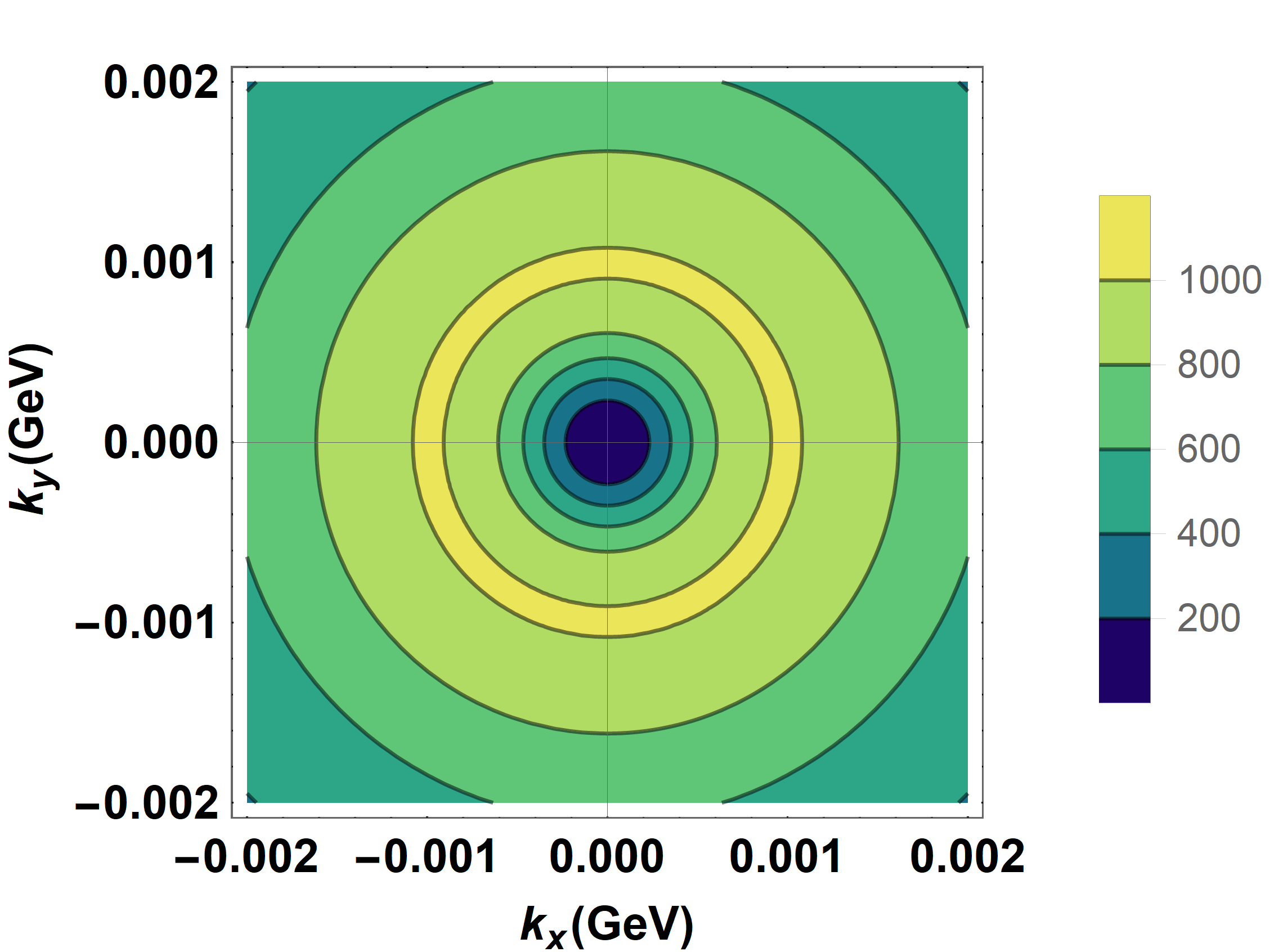} }}%
    \qquad
    \subfloat[\centering $x\rho^{\circlearrowleft/\leftrightarrow}_g (x,k_\perp) $]{{\includegraphics[scale=0.68]{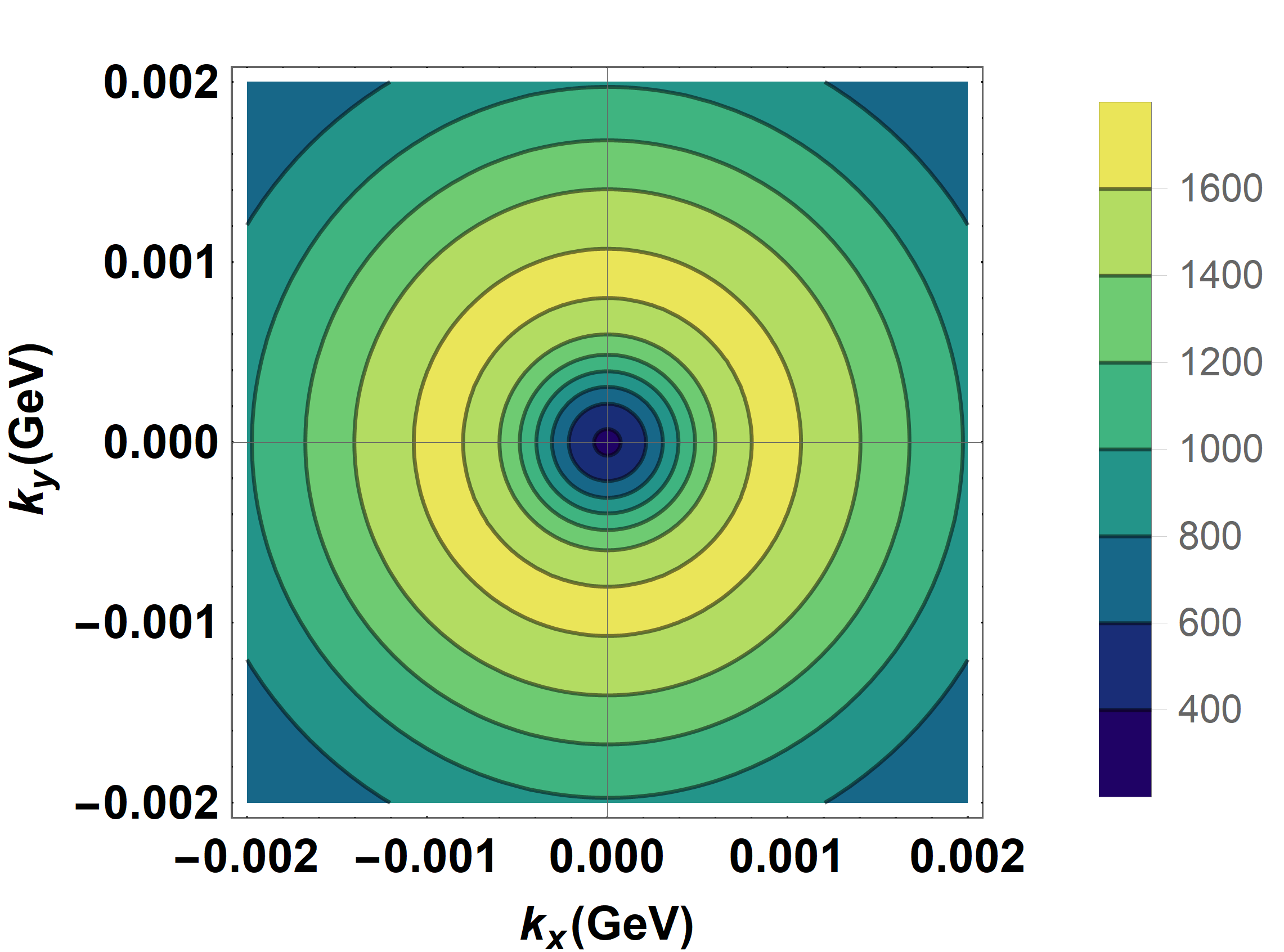} }}%
    \caption{Plots of gluon (a) helicity density and (b) worm-gear density in momentum space for $x=0.3$. Mass of the target $m=3.3$ MeV for both plots.}
    \label{fig: helicity and worm gear plots}
\end{figure}

 The helicity density $x\rho^{\circlearrowleft/+}_g(x,k_\perp)$, and worm-gear density $x\rho^{\circlearrowleft/\leftrightarrow}_g (x,k_\perp) $ are defined as \cite{Bacchetta:2020vty}
 \begin{align}
    x\rho^{\circlearrowleft/+}_g(x,k_\perp)&=xf_1^g(x,k_\perp^2)+xg_{1L}^g(x,k_\perp^2),\label{eq: helicity density}\\
    x\rho^{\circlearrowleft/\leftrightarrow}_g (x,k_\perp)&=xf_1^g(x,k_\perp^2)-\frac{k_x}{m}xg_{1T}^g(x,k_\perp^2).\label{eq: worm-gear density}
\end{align}
 Helicity density describes the probability density of  circularly polarized gluons with given $x$ and $k_\perp$ inside the longitudinally polarized target, and worm-gear density describes the probability density of circularly polarized gluons with longitudinal momentum fraction $x$ and transverse momentum $k_\perp$ inside the transversely polarized target. The analytical results of Eqs(\ref{eq: helicity density},\ref{eq: worm-gear density}) are presented as a contour plot in the transverse momentum space in Fig \ref{fig: helicity and worm gear plots}.
The left plot indicates helicity distributions and the right plot is for worm gear density for gluon. In the light-front spectator mode model \cite{Chakrabarti:2023djs}, the gluon helicity distributions demonstrated cylindrical symmetry along the direction of the proton's motion ($P^+$ direction), while the worm gear density showed an asymmetric distribution around the longitudinal direction of motion. In contrast, in our model, both distributions exhibit cylindrical symmetry around the direction of motion of the target. Nevertheless, the magnitude of gluon densities in momentum space have qualitative agreement with the findings in \cite{Chakrabarti:2023djs,Bacchetta:2020vty}, showing minimal density for the zero momentum region and oscillations as we move away from the origin.
%
\subsection{Spin-Spin Correlation}
\begin{figure}[htp!]
\begin{minipage}[c]{1\textwidth}
\small{(a)}\includegraphics[width=7.9cm,height=7cm,clip]{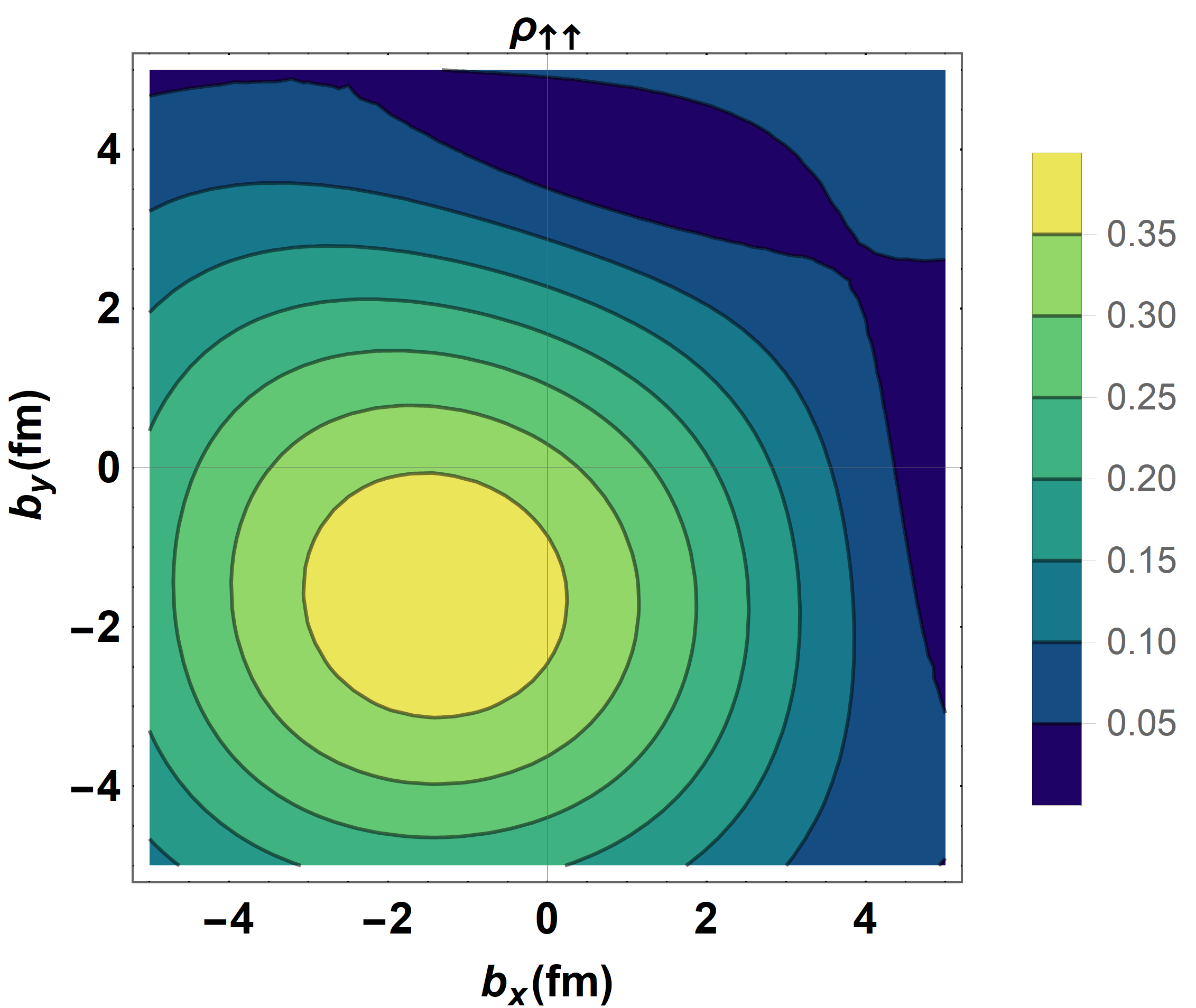}
\small{(b)}\includegraphics[width=7.9cm,height=7cm,clip]{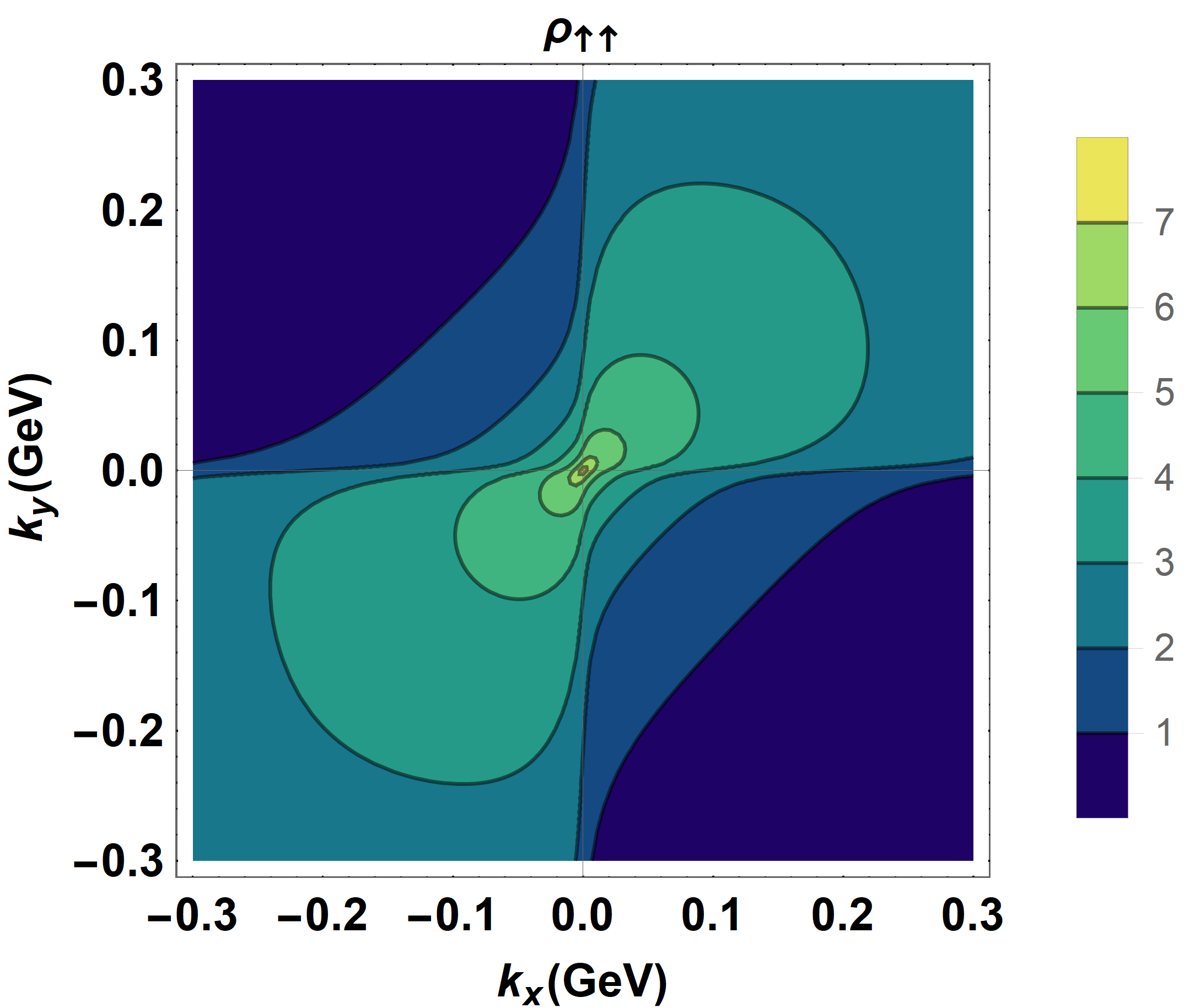}
\small{(c)}\includegraphics[width=7.9cm,height=7cm,clip]{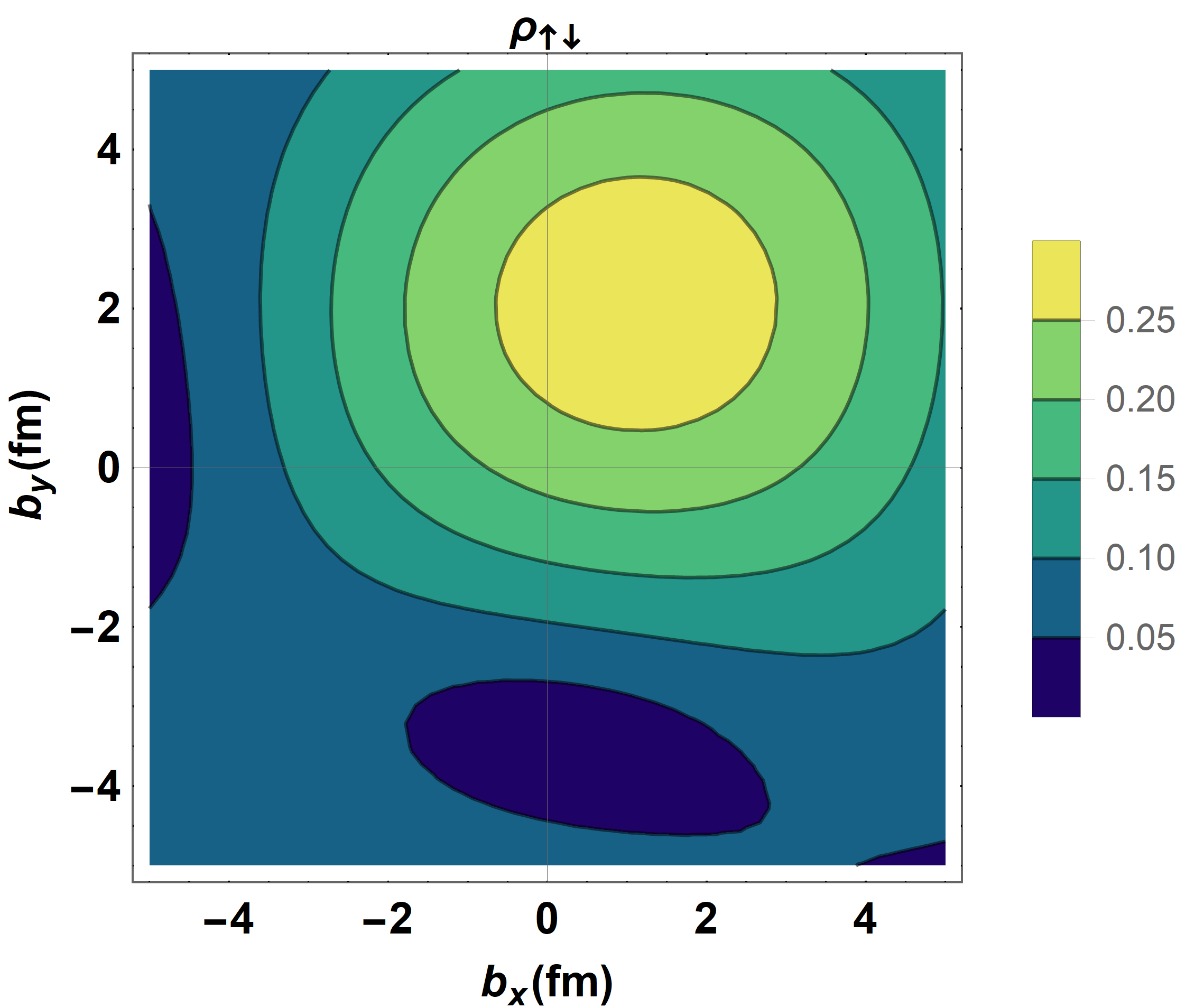}
\small{(d)}\includegraphics[width=7.9cm,height=7cm,clip]{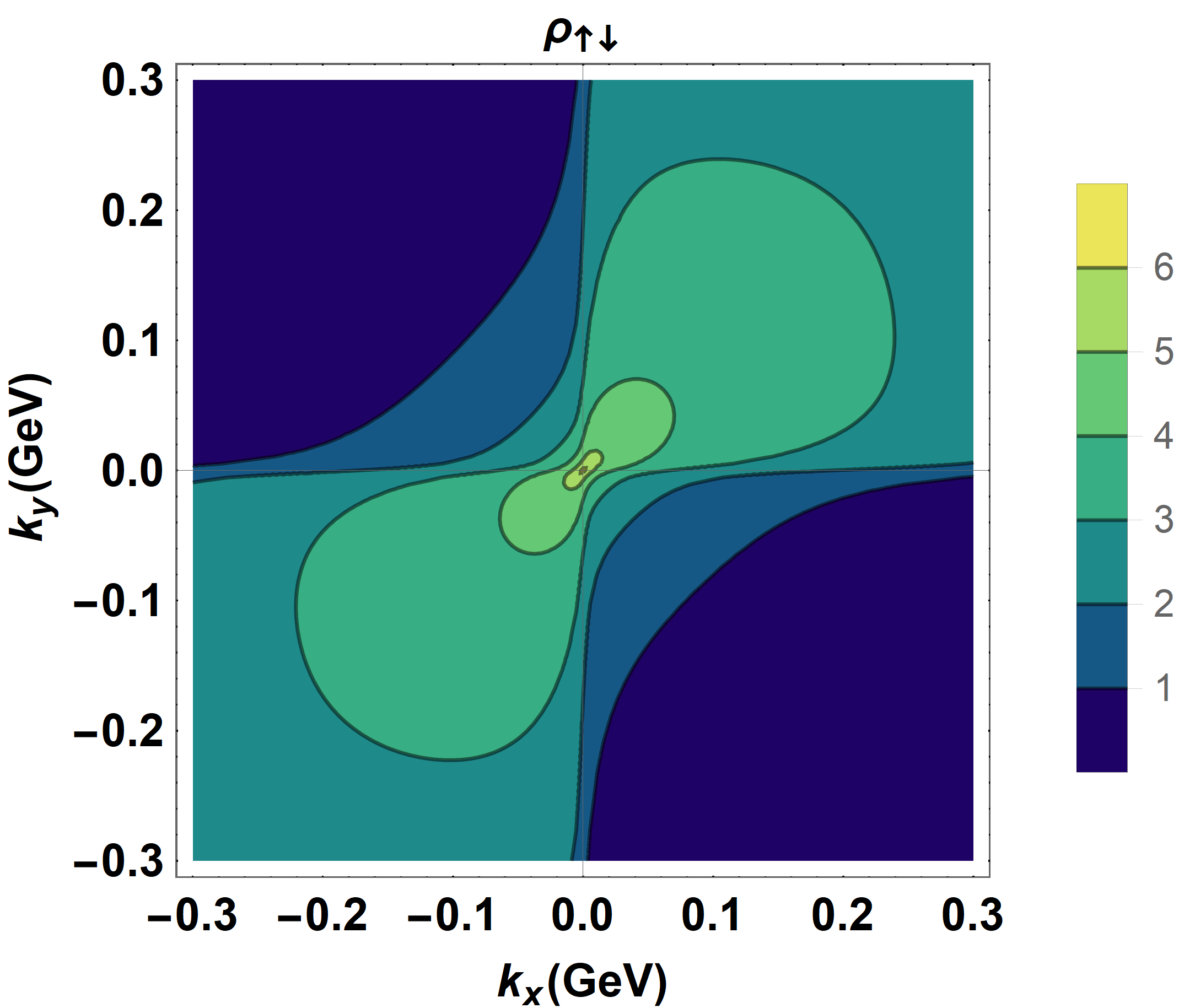}
\end{minipage}
\caption{\label{fig: gluon density position, momentum space}The distribution $\rho_{\Lambda\lambda}$ (Eq.(\ref{Eq:spin-spin correlations})) in position and momentum space for $x=0.3$ when the polarization of gluon and target state are aligned in same direction (a),(b), and aligned in opposite direction (c),(d). We choose $k_\perp=0.4\hat{y}$ for b-space, $b_\perp=0.4\hat{y}$ for k-space plots, and mass of the target state $m$ as $3.3$ MeV.}
\end{figure}
The spin-spin correlations can be investigated by tacking the combination of all the helicity combinations having corresponding Wigner distributions as weight factors. The Wigner distribution for gluon with the longitudinal polarization $\lambda$ in a strongly bound state with the longitudinal polarization $\Lambda$, can be defined as \cite{Lorce:2011kd}
\begin{align}
\rho_{\Lambda\lambda}(\Vec{b_\perp},\Vec{k_\perp},x)=\frac{1}{2}\Big[\rho_{UU}(\Vec{b_\perp},\Vec{k_\perp},x)+\Lambda\rho_{LU}(\Vec{b_\perp},\Vec{k_\perp},x)+\lambda\rho_{UL}(\Vec{b_\perp},\Vec{k_\perp},x)+\Lambda\lambda\rho_{LL}(\Vec{b_\perp},\Vec{k_\perp},x)\Big], \label{Eq:spin-spin correlations}
\end{align}
where $\rho_{UU}$ is the Wigner distribution of unpolarized gluon in unpolarazied target, and other three parts $\rho_{LU}$, $\rho_{UL}$, $\rho_{LL}$ gives the distortion due to longitudinal polarization of gluon or target state.

Examining the contour plots of Wigner distribution $\rho_{\Lambda\lambda}$ in both position and momentum space offers intriguing insights into the correlation between gluon and target state spin. In Fig\ref{fig: gluon density position, momentum space}, we present the contour plot of the Wigner distribution in position and momentum space, illustrating cases where the spin of the gluon and target state is aligned in the same direction (4(a), 4(b)) and in opposite directions (4(c), 4(d)). Notably, we observe a diagonal shift in the entire distribution when there is a spin reversal in either the gluon or the target state. It is worth mentioning that while a horizontal shift in the distribution for quarks due to spin flip has been noted in other models \cite{Lorce:2011kd,Chakrabarti:2017teq}, our model introduces a distinctive diagonal shift for the gluon. However, no observable shift is noticed in the Wigner distribution of gluons in momentum space, suggesting its independence from the spin flip of the gluon or target state. This phenomenon sets our model apart from others, where shifts, if observed, are typically horizontal.

\begin{figure}[htp!]
\begin{minipage}[c]{1\textwidth}
\small{(a)}\includegraphics[width=7.9cm,height=7cm,clip]{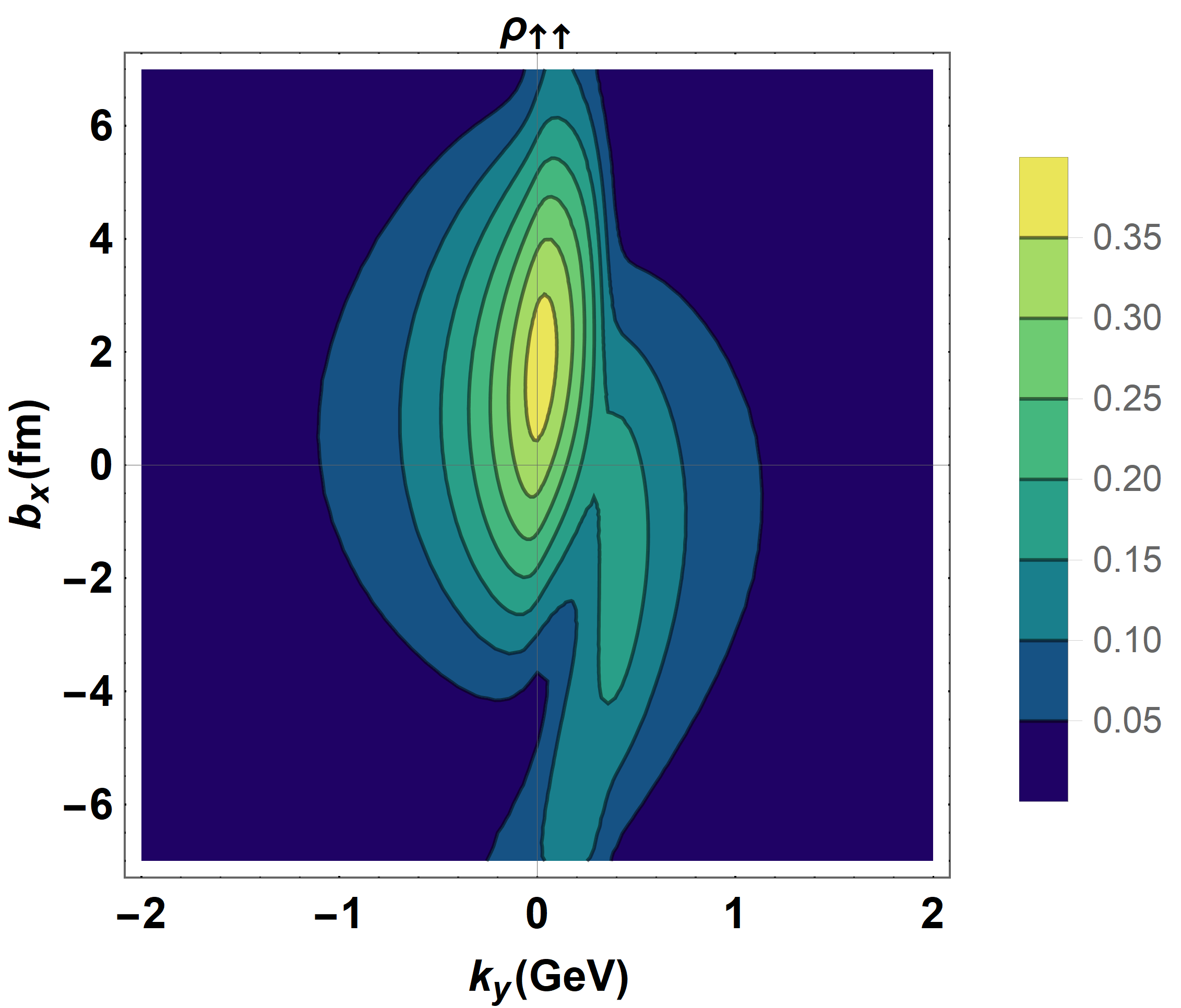}
\small{(b)}\includegraphics[width=7.9cm,height=7cm,clip]{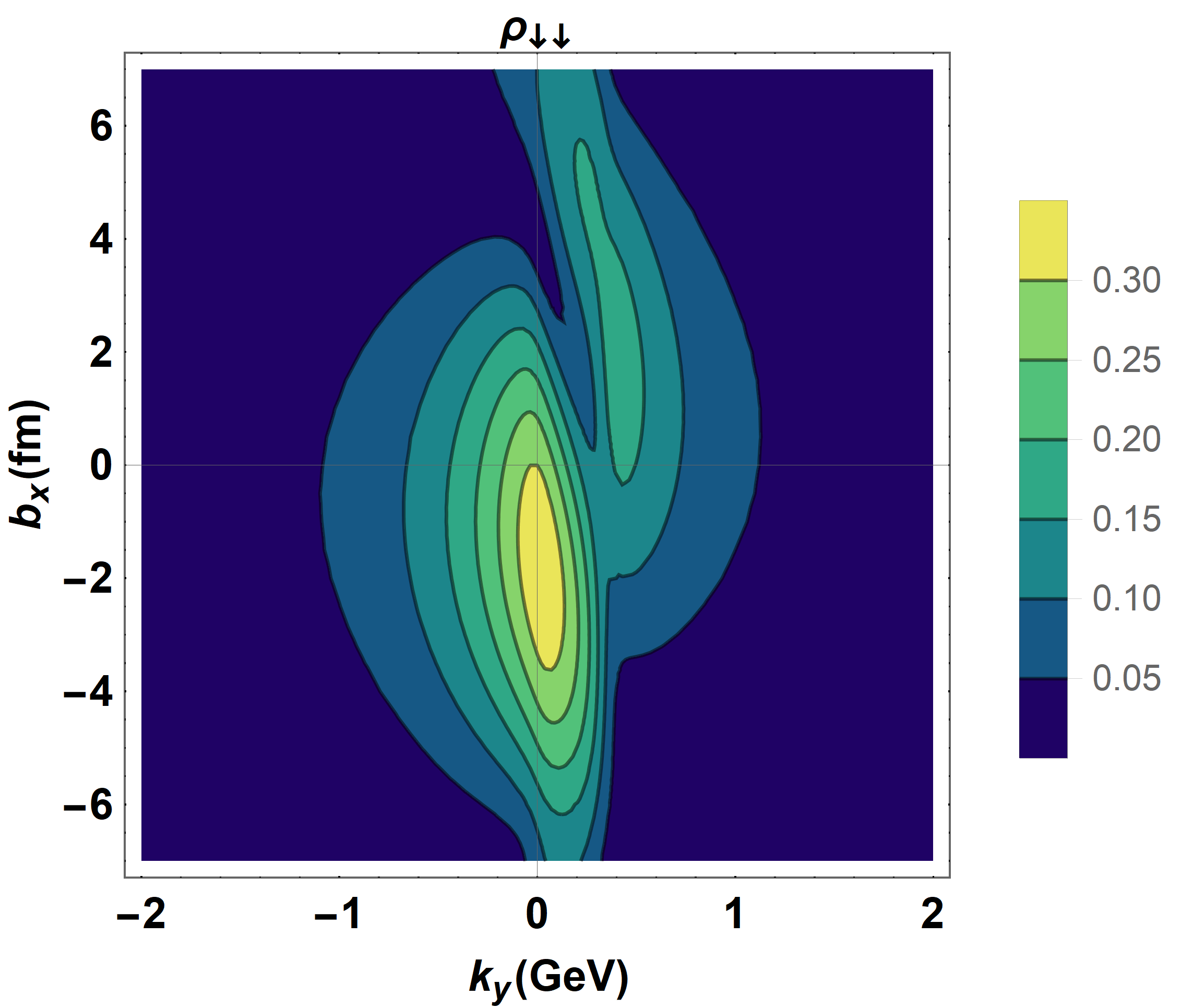}
\small{(c)}\includegraphics[width=7.9cm,height=7cm,clip]{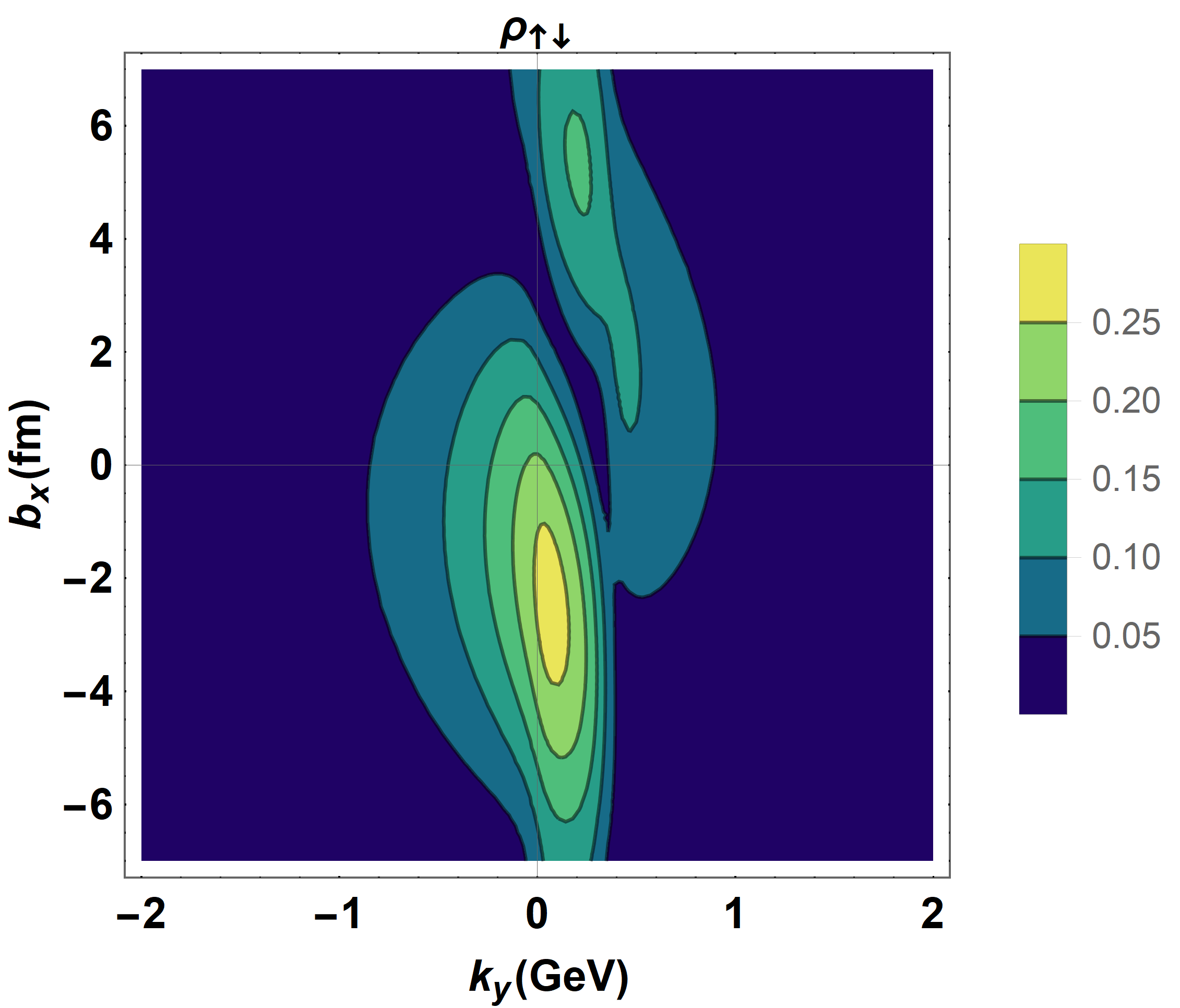}
\end{minipage}
\caption{\label{fig: gluon density mixed space}The distribution $\rho_{\Lambda\lambda}$ (Eq.(\ref{Eq:spin-spin correlations})) in mixed space for $x=0.3$ when the polarization of gluon and target state are aligned in same direction ((a),(b)), and aligned in opposite direction (c). We choose $k_x=0.4$ GeV and $b_y=0.4$ fm, and took the mass of target state $m$ as $3.3$ MeV for all plots.}
\end{figure}

In Fig \ref{fig: gluon density mixed space}, the distribution in mixed space reveals a dipole structure. The dipole becomes more defined and separates as the spin of the gluon and target aligns in opposite directions. This unique feature adds depth to our understanding of the interplay between spin orientations and spatial distributions in the context of gluon and target states.
%
%

%

\newpage

\section{Wigner distribution in boost invariant longitudinal space\label{Sec_WD_sigma}}
The Wigner distributions are defined by the Fourier transformation of GTMDs and contain the non-perturbative information on momentum space $(x,k_\perp)$ as well as on position equivalent spaces ($\sigma, b_\perp$). The transverse impact parameter $b_\perp$ is Fourier conjugate of the transverse momentum transfer $D_\perp=\Delta_\perp/(1-\xi^2)$ from initial to final state \cite{Diehl:2002he,Burkardt:2002hr,Ralston:2001xs,Kaur:2018ewq}.  The Wigner distribution in the transverse impact parameter space $W^{ [\Gamma]}_{[\lambda\lambda^{\prime}]}(x_g,\xi,b_\perp, k_\perp)$ is defined by the Fourier transformation of GTMDs correlator of Eq.(\ref{qqc}) over $D_\perp$. For gluon, the WDs in transverse impact parameter space $W^{ [\Gamma]}_{[\lambda\lambda^{\prime}]}(x_g,\xi,b_\perp, k_\perp)$  in the dressed quark model, is discussed for $\xi=0$ limit \cite{more2018wigner,mukherjee2015wigner}.

In this section, we present the Wigner distribution in the boost invariant longitudinal space $\sigma$, which is the Fourier conjugate of longitudinal momentum transfer skewness $\xi$ and defined as  $\sigma=\frac{1}{2}b^- P^+$ \cite{Brodsky:2006in,Brodsky:2006ku}. 
The Wigner distribution for gluon in longitudinal impact parameter ($\sigma$) space is defined as 
\begin{align} \label{rho_sigma}
 \rho^{[\Gamma]}(x,\sigma,\Delta_\perp,k_\perp;S)&=\int_{0}^{\xi_{max}}\frac{d\xi}{2\pi}e^{i\sigma\cdot\xi}W^{[\Gamma]}(x,\xi,\Delta_\perp,k_\perp;S)
\end{align}
 where GTMDs correlator is the same to Eq.(\ref{qqc}) having the same helicity ($S$) of the final and initial states. The upper limit of the integration is restricted by the energy transfer $t$ to the system as 
         \begin{equation}
          \xi_{max}=\frac{-t}{2m^2}\Big(\sqrt{1+\frac{4m^2}{-t}}-1\Big) ~;~~~~
         \text{and}~~-t=\frac{4\xi^2m^2+\Delta^2_\perp}{1-\xi^2}. \label{rel_t_Del}
        \end{equation}

 To investigate the different polarization combinations of the dressed state and the gluon, we define $\rho^g_{X,Y}$, where $X$ and $Y$ represent the polarization of the dressed quark state and gluon, respectively. The subscripts $X, Y = U, L, T $ for the unpolarized, longitudinally polarization and transverse polarization respectively. For the different polarization combinations of $X$ and $Y$, the Wigner distribution for gluon is defined as
\be
\rho^g_{UY}(x_g,\sigma,\Delta_\perp,k_\perp) &=& \frac{1}{2}\Big[\rho^{[\Gamma]}(x_g,\sigma,\Delta_\perp,k_\perp,+\hat{e}_z) +
\rho^{[\Gamma]}(x_g,\sigma,\Delta_\perp,k_\perp,-\hat{e}_z) \Big] \label{rUY}\\
\rho^g_{LY}(x_g,\sigma,\Delta_\perp,k_\perp) &=& \frac{1}{2}\Big[\rho^{[\Gamma]}(x_g,\sigma,\Delta_\perp,k_\perp,+\hat{e}_z) -
\rho^{[\Gamma]}(x_g,\sigma,\Delta_\perp,k_\perp,-\hat{e}_z) \Big] \label{rLY} 
\ee
The above equations, Eqs.(\ref{rUY},\ref{rLY}), are the composite form of four Wigner distributions corresponding to the two different polarization of gluon $Y=\{U, L\}$. Gamma in the superscript can have  $\Gamma = \{\delta^{ij},-i\epsilon^{ij}_\perp\}$ for unpolarized and longitudinally polarized gluons respectively. 

The model result for the Wigner distributions corresponding to the unpolarized dressed quark state $\rho^g_{UU}, \rho^g_{UL}$ and the longitudinally polarized target state $\rho^g_{LU}, \rho^g_{LL}$ are given by,
           \begin{align} 
                \rho^g_{UU}(x,\sigma,\Delta_\perp,k_\perp)&=\int_{0}^{\xi_{max}}\frac{d\xi}{2\pi}e^{i\sigma\cdot\xi}\sqrt{1-\xi^2}S^{0,+}_{1,ia} \label{rho_uu} \\
     \rho^g_{LL}(x,\sigma,\Delta_\perp,k_\perp)&=\int_{0}^{\xi_{max}}\frac{d\xi}{2\pi}e^{i\sigma\cdot\xi}\sqrt{1-\xi^2}S^{0,-}_{1,ia}\label{r_TTp}\\
     \rho_{LU}^g(x,\sigma,\Delta_\perp,k_\perp)&=\int_{0}^{\xi_{max}}\frac{d\xi}{2\pi}e^{i\sigma\cdot\xi}\frac{i(k_1\Delta_2-k_2\Delta_1)S^{0,-}_{1,ib}\sqrt{1-\xi^2}}{m^2}\\
        \rho_{UL}^g (x,\sigma,\Delta_\perp,k_\perp)&=\int_{0}^{\xi_{max}}\frac{d\xi}{2\pi}e^{i\sigma\cdot\xi}\frac{i(k_1\Delta_2-k_2\Delta_1)S^{0,+}_{1,ib}\sqrt{1-\xi^2}}{m^2} \label{rho_uL}
\end{align}
          
 \begin{figure}[!htp]
\begin{minipage}[c]{1\textwidth}
\small{(a)}\includegraphics[width=7.5cm,height=5.5cm,clip]{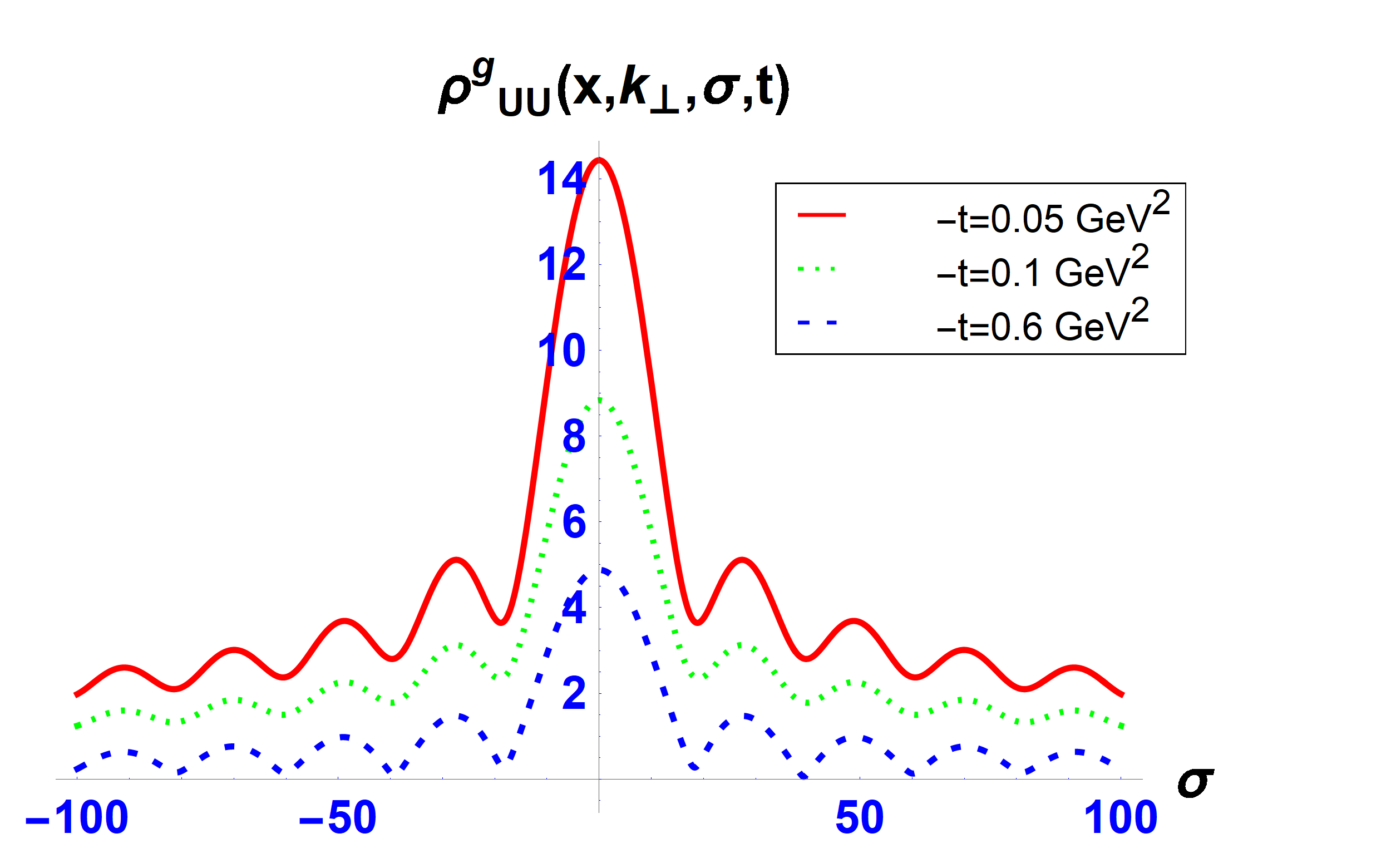}
\hspace{0.1cm}
\small{(b)}\includegraphics[width=7.5cm,height=5.5cm,clip]{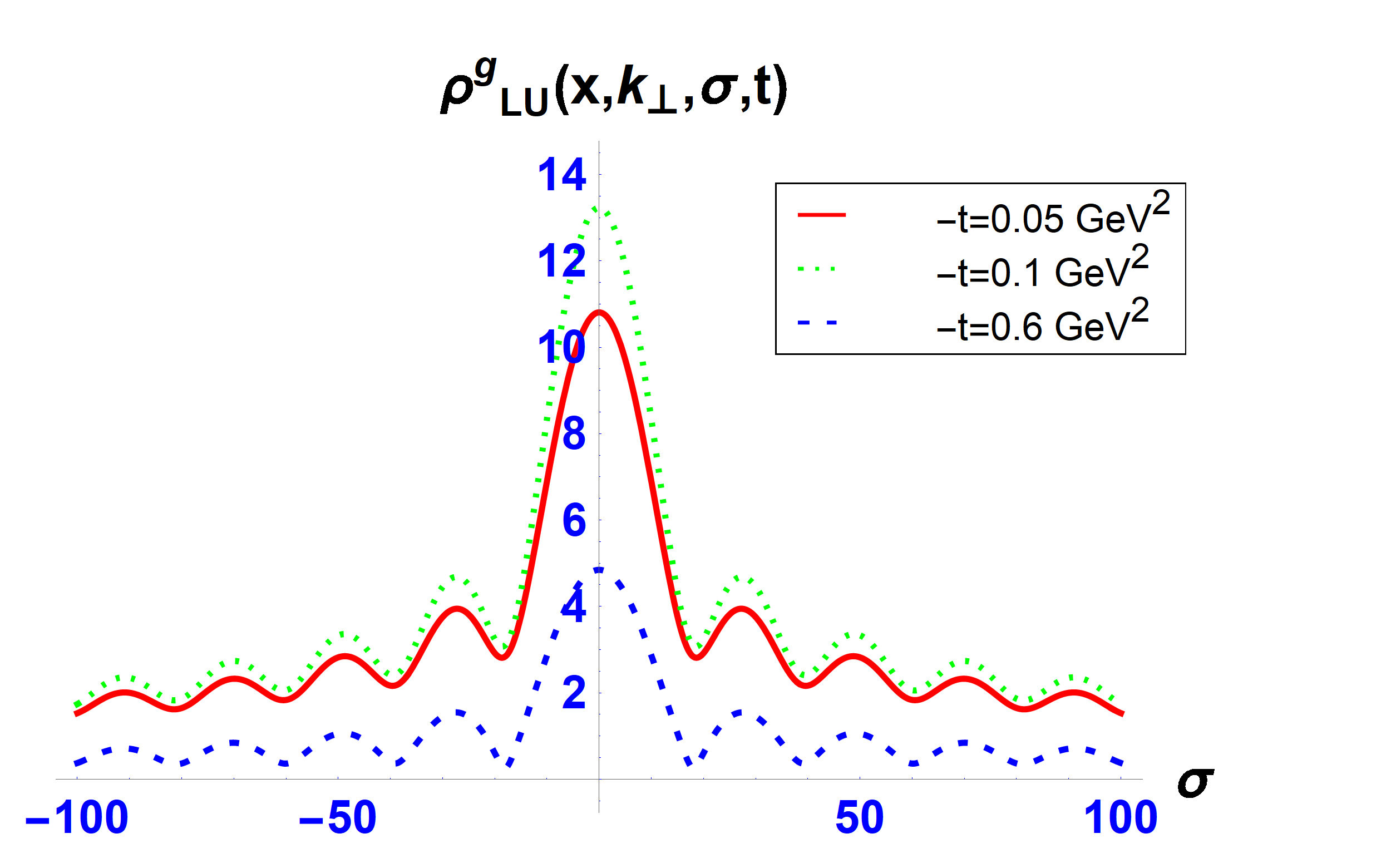} \\
\small{(c)}\includegraphics[width=7.5cm,height=5.5cm,clip]{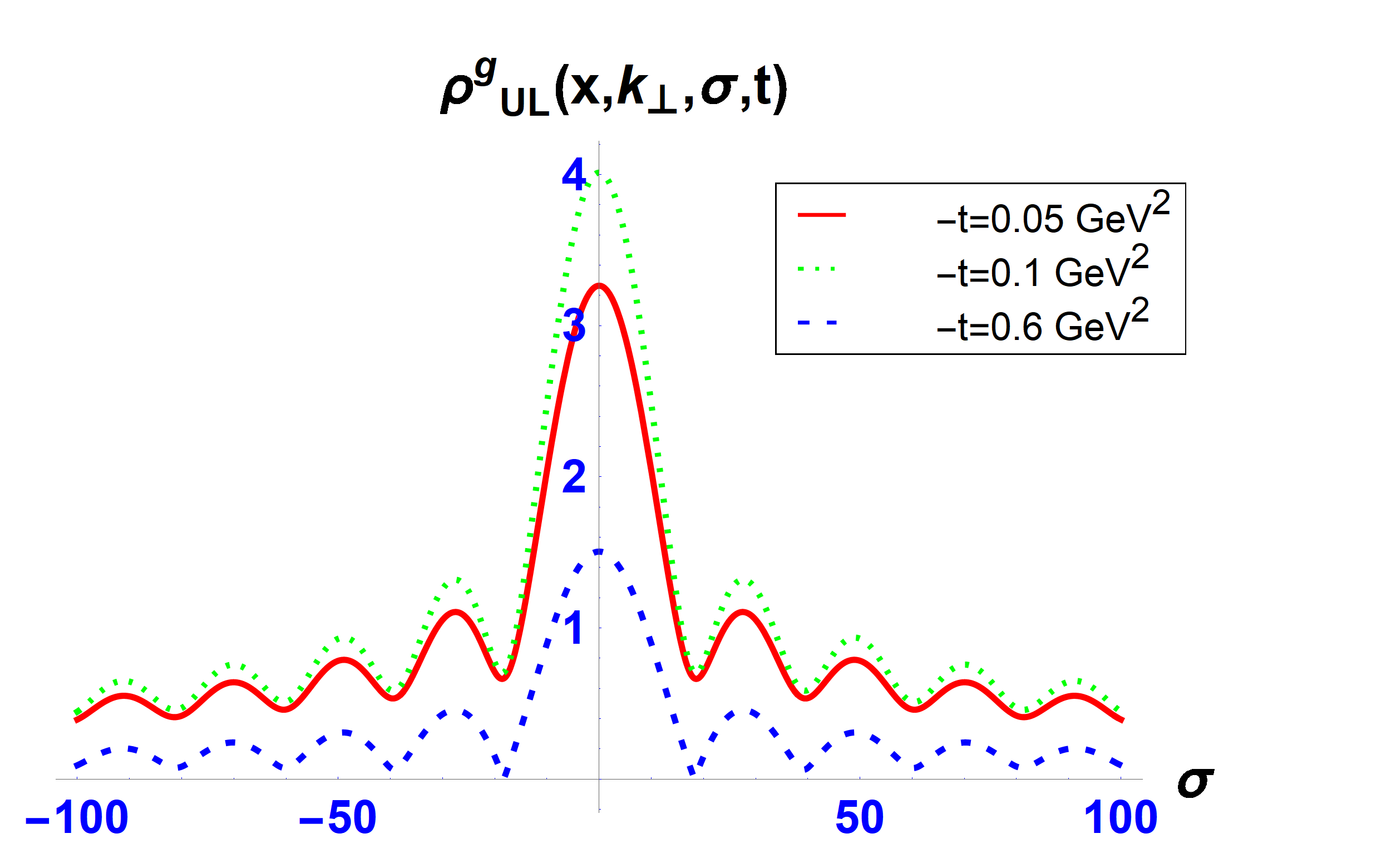}
\hspace{0.1cm}
\small{(d)}\includegraphics[width=7.5cm,height=5.5cm,clip]{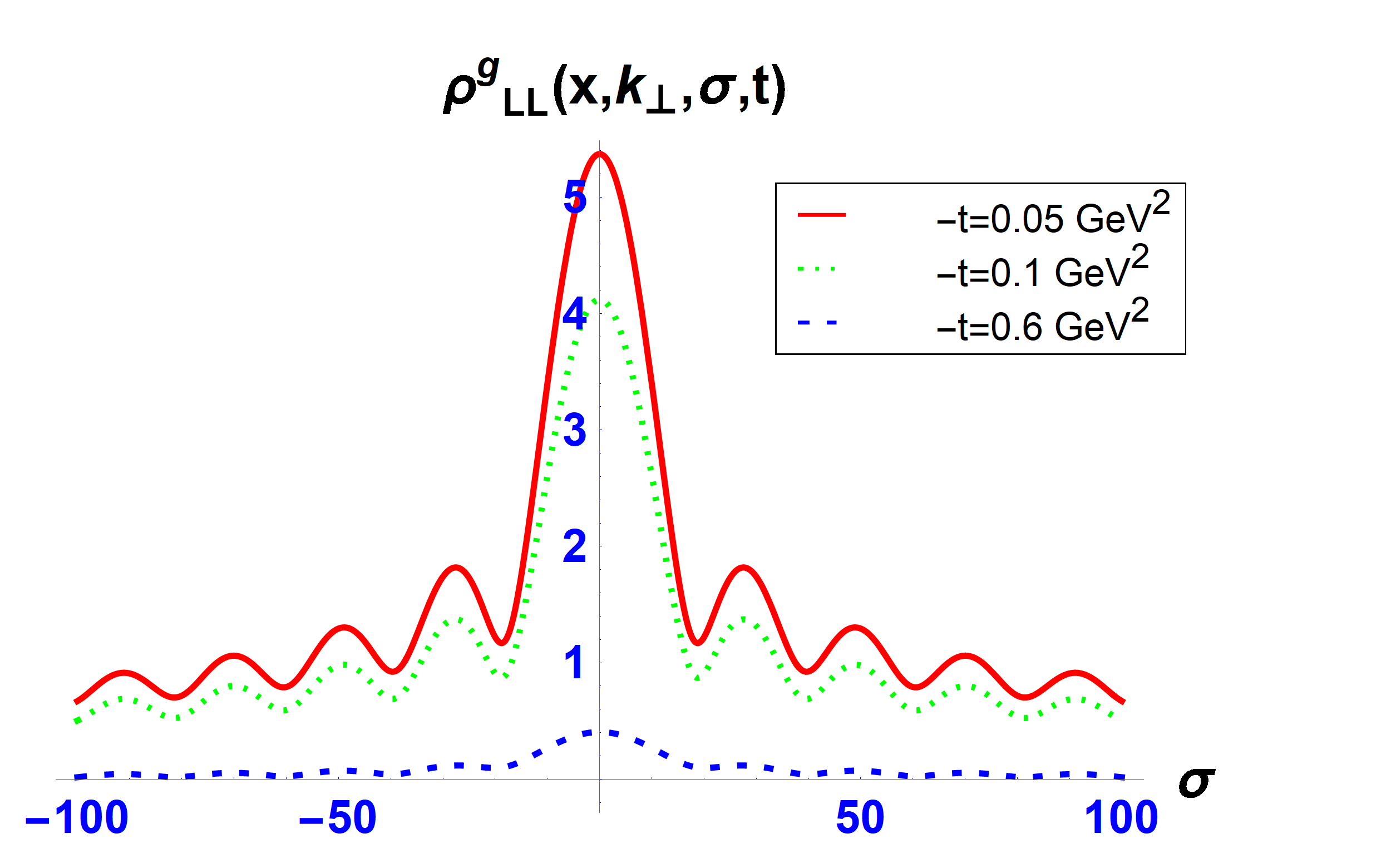} \\
\end{minipage}
\caption{\label{fig_sig}
 Gluon Wigner distribution (in $\sigma$-space) at different values of $-t$ for (a) unpolarized gluon in unpolarized target (b) unpolarized gluon in longitudinally polarized target (c) longitudinally polarized gluon in unpolarized target (d) longitudinally polarized gluon in longitudinally polarized target.}
\end{figure}

Using the model result of GTMDs listed in Eqs.(\ref{gtmd-unp-1}-\ref{gtmd-lnp-4}), the Fourier transformation is performed over skewness ($\xi$) as defined in Eq.(\ref{rho_sigma})
and end up with the function of  $\sigma$.
We present the $\sigma$-space variation of these Wigner distributions (Eq.(\ref{rho_uu}-\ref{rho_uL})) for fixed value of $x=0.3$ in Fig.(\ref{fig_sig}). 
At the time of integration, we have replaced total transverse momentum transfer ($\Delta_\perp$) by $t$ using the relation given in Eq.(\ref{rel_t_Del}), and consider $\Delta_\perp \cdot k_\perp=0 $. The Wigner distributions are shown for the range of $\sigma=\{-100 ~ {\rm to} ~100\}$. The three labels with red continuous, green dotted and blue dashed lines correspond to the different values of $-t =(0.05, 0.1, 0.6) ~GeV^2$. The mass of the quark is fixed at 0.0033 GeV. 

  This model result shows oscillatory behaviors that have qualitative similarity to the diffraction pattern of single-slit optical phenomenon. Here $\sigma$ is comparable to the distance of the slit from the screen and skewness ($\xi$) plays the role of slit-width. We observed that the width and peak of the central maxima reduced with the increasing value of $-t$. A similar oscillatory pattern is found for the quark sector of this model. Diffraction-analogous behavior is also reported in other models e.g, Light front quark diquark model\cite{Maji:2022tog}, holographic QCD\cite{Brodsky:2006ku} etc., and our result shows similar qualitative behavior.

  \section{Conclusion}

In our prior publication \cite{Ojha:2022fls}, we reported the GTMDs related to the quark sector within the dressed quark model for non-zero skewness. In the present study, we extend our analysis to the gluon sector, focusing on the computation of eight GTMDs with non-zero skewness. Specifically, we explore scenarios involving unpolarized and longitudinally polarized gluons within a dressed quark system. The calculations for gluon-gluon correlators to extract the GTMDs involve the utilization of the two-particle light-front wave function within our model.

The derived analytical expression for GTMDs with non-zero skewness successfully adheres to the essential hermiticity constraint. Similar to quark GTMDs, the T-odd components of gluon GTMDs vanish, rendering the GTMDs real. 
The obtained expressions for GTMDs are subsequently employed to investigate various properties of the gluon in the dressed quark system. These explorations encompass aspects such as orbital angular momentum, spin-orbit correlation, transverse momentum-dependent parton distributions (TMDs), helicity density, worm-gear density, and spin-spin correlation.

Notably, this work introduces, for the first time, the gluon Wigner distribution in the boost-invariant longitudinal space. We present the variation of the gluon Wigner distribution in the boost-invariant longitudinal $\sigma$ space, focusing on different values of the variable $t$ while maintaining $x$ at a fixed value of $0.3$. Our model's results reveal an oscillatory pattern reminiscent of diffraction patterns observed in optical single-slit experiments. The regulation factor influencing the intensity of the peak is the energy transfer to the system, denoted as $-t$. As $-t$ increases, the peak of the Wigner distribution in the $\sigma$ space diminishes. Similar diffraction
patterns are identified for the quark sectors in our model \cite{Ojha:2022fls} and in other models \cite{Maji:2022tog, Brodsky:2006ku}.
A potential avenue for future work in this direction involves the assessment of GTMDs and Wigner distribution in the σ space for transversely polarized gluons.

 \section*{Acknowledgment}
V.K.O. is supported by the seed grant project, Sardar Vallabhbhai National Institute of Technology (SVNIT) Surat, with the assigned project number 2021-22/DOP/05. The authors are thankful to Dipankar Chakrabarti for the useful discussions. TM is supported by the Science and Engineering Research Board (SERB) through the SRG (Start-up Research Grant) of File No. SRG/2023/001093.

 \newpage
\appendix

\section{GTMDs CALCULATIONS}
\begin{itemize}
\item We use the following definition of Dirac spinor 
\begin{align}
    u_\uparrow(p)=\frac{1}{\sqrt{2p^+}}\begin{pmatrix}
    p^++m\\
    p^1+ip^2\\
    p^+-m\\
    p^1+ip^2
    \end{pmatrix}\;
    \text{and}\;\;
  u_\downarrow(p)=\frac{1}{\sqrt{2p^+}}\begin{pmatrix}
  -p^1+ip^2\\
    p^++m\\
    p^1-ip^2\\
    -p^++m
    \end{pmatrix}   
\end{align}
 \item Kinematics for quark and gluon:
    \begin{table}[h!]
    \centering{}
    \small \begin{tabular}{| c | c | c|}
    \hline
        Kinematic Variables &  Quark & Gluon \\
       \hline\hline
         Average initial longitudinal momentum fraction & $x'=\frac{x-\xi}{1-\xi}$   & $x'_g=1-\frac{x-\xi}{1-\xi}=\frac{x_g}{1-\xi}$ \\
       \hline
       Average final longitudinal momentum fraction & $y=\frac{\xi+x}{\xi+1}$ & $y_g=1-\frac{x+\xi}{1+\xi}=\frac{x_g}{1+\xi}$ \\
       \hline
          Initial transverse Jacobi momentum & $q'_\perp=k_\perp+\frac{(1-x)\Delta_\perp}{2(1-\xi)}$  & $q'_{\perp g}=-k_\perp-\frac{(1-x)\Delta_\perp}{2(1-\xi)}=-k_\perp-\frac{x_g\Delta_\perp}{2(1-\xi)}$ \\
      \hline
      Final transverse Jacobi momentum & $q_\perp=k_\perp-\frac{(1-x)\Delta_\perp}{2(1+\xi)}$  & $q_{\perp g}=-k_\perp+\frac{(1-x)\Delta_\perp}{2(1+\xi)}=-k_\perp+\frac{x_g\Delta_\perp}{2(1+\xi)}$   \\
      \hline

    \end{tabular}
    \caption{ A comparison of kinematics for quark and gluon. Quark is the struck parton.}
    \label{tab2}
\end{table}
 \item Eq.(\ref{ggc-unpolarized}) gives the following for different combination of polarization:
\small
  \begin{align}
W^{(\delta^{ij}_\perp)}_{++}+W^{(\delta^{ij}_\perp)}_{--}&=N\frac{(4(1-\xi^2)k^2_\perp-(1-x)^2\Delta^2_\perp+4\xi(1-x)k_\perp\cdot\Delta_\perp)((1-x_g)^2-\xi+1)+4m^2(x^2-\xi^2)^2}{D^*(q'_{\perp g},x'_g)D(q_{\perp g},y_g)\sqrt{\frac{x^2-\xi^2}{1-\xi^2}}(\xi^2-x^2)(1-x)^2}\label{app:ggc1}\\
W^{(\delta^{ij}_\perp)}_{++}-W^{(\delta^{ij}_\perp)}_{--}&=N\frac{-4i(x^2-2x+\xi^2)(k_2\Delta_1-k_1\Delta_2)}{D^*(q'_{\perp g},x'_g)D(q_{\perp g},y_g)\sqrt{\frac{x^2-\xi^2}{1-\xi^2}}(x^2-\xi^2)(1-x)}\label{app:ggc2}\\
W^{(\delta^{ij}_\perp)}_{+-}+W^{(\delta^{ij}_\perp)}_{-+}&=N\frac{4im(2\xi(x^2-2x+\xi^2)k_1-(1-x)(x^2+\xi^2)\Delta_1)}{D^*(q'_{\perp g},x'_g)D(q_{\perp g},y_g)\sqrt{\frac{x^2-\xi^2}{1-\xi^2}}(x^2-\xi^2)(1-x)}\label{app:ggc3}\\
W^{(\delta^{ij}_\perp)}_{+-}-W^{(\delta^{ij}_\perp)}_{-+}&=N\frac{-4m(2\xi(x^2-2x+\xi^2)k_2-(1-x)(x^2+\xi^2)\Delta_2)}{D^*(q'_{\perp g},x'_g)D(q_{\perp g},y_g)\sqrt{\frac{x^2-\xi^2}{1-\xi^2}}(x^2-\xi^2)(1-x)}\label{app:ggc4}
\end{align}
\item Eqs.(\ref{bilinear decomposition unpol}) leads to the following: 
\begin{align}
    W^{(\delta^{ij}_\perp)}_{++}+W^{(\delta^{ij}_\perp)}_{--}&=2S^{0,+}_{1,ia}\label{app-gtmd1}\\
    W^{(\delta^{ij}_\perp)}_{++}-W^{(\delta^{ij}_\perp)}_{--}&=-\frac{2iS^{0,+}_{1,ib}(k_2\Delta_1-k_1\Delta_2)}{m^2}\label{app-gtmd2}\\
    W^{(\delta^{ij}_\perp)}_{+-}+W^{(\delta^{ij}_\perp)}_{-+}&=\frac{2i(k_2P^{0,+}_{1,ia}+\Delta_2P^{0,+}_{1,ib})}{m}\label{app-gtmd3}\\
    W^{(\delta^{ij}_\perp)}_{+-}-W^{(\delta^{ij}_\perp)}_{-+}&=\frac{2(k_1P^{0,+}_{1,ia}+\Delta_1 P^{0,+}_{1,ib})}{m}\label{app-gtmd4}
\end{align}
 GTMDs $S^{0,+}_{1,ia},S^{0,+}_{1,ib},P^{0,+}_{1,ia}$, $P^{0,+}_{1,ib}$ can be obtained using Eqs.(\ref{app:ggc1}-\ref{app:ggc4}) and Eqs.(\ref{app-gtmd1}-\ref{app-gtmd4}). A similar procedure for $\Gamma=-i\epsilon^{ij}_\perp$ gives the GTMDs $S^{0,-}_{1,ia},\;S^{0,-}_{1,ib},\;P^{0,-}_{1,ia},\;P^{0,-}_{1,ib}$.
\end{itemize}

\section{Gluon GTMDs in the notation of \cite{Meissner:2009ww} \label{appc} }
\begin{itemize}
    \item For unpolarized gluon
    \begin{align}
  F_{11}^g=&-\frac{\alpha_g\sqrt{1-\xi^2}}{2x_g}\Big[4(1-\xi^2)(1+x_g^2-\xi^2)k_\perp^2-x_g^2(1+x_g^2-\xi^2)\Delta^2_\perp+4x_g\xi(1+x_g^2-\xi^2)k_\perp\cdot\Delta_\perp\nn\\
  +& 4m^2((1-x_g)^2-\xi^2)^2\Big] \label{eq: gluon F11}\\
  F_{12}^g=& \frac{1}{\sqrt{1-\xi^2}}\Big[-(2m^2x_g\beta_g((1-x_g)^2+\xi^2)+\xi\alpha_g(x_g^2+\xi^2-1))\Delta^2_\perp+\nn\\
  & 4m^2\beta_g\xi(x_g^2+\xi^2-1)k_\perp\cdot\Delta_\perp\Big]\\
  F_{13}^g=& \frac{1}{4x_g\sqrt{1-\xi^2}}\Big[-4(4\beta_gx_gm^2\xi(x_g^2+\xi^2-1)+\alpha_g(1+x_g^2-\xi^2)(1-\xi^2))k_\perp^2\nn\\
  +&\alpha_gx_g^2(1+x_g^2-\xi^2)\Delta^2_\perp
  + 4x_g(2\beta_gx_gm^2((1-x_g)^2+\xi^2)-\alpha_g\xi(1-(x_g^2+\xi^2-1)))k_\perp\cdot\Delta_\perp \nn\\
  -& 4m^2\alpha_g((1-x_g)^2-\xi^2)^2\Big]\\
  F_{14}^g=& 2m^2\alpha_g\sqrt{1-\xi^2}(x_g^2+\xi^2-1)
    \end{align}
\item For longitudinally polarized gluon
 \begin{align}
         G_{11}^g=&2\alpha_gm^2\sqrt{1-\xi^2}(x_g^2-\xi^2+1)\\
G_{12}^g=&\frac{1}{\sqrt{1-\xi^2}}\Big[-(4m^2\beta_gx_g\xi(1-x_g)+\alpha_g(x_g^2-\xi^2+1))\Delta_\perp^2-4m^2\beta_g((1-x_g)^2-\nn\\
&\xi^2(1-2x_g))k_\perp\cdot\Delta_\perp\Big]\\
G_{13}^g=&\frac{1}{4x_g\sqrt{1-\xi^2}}\Big[4(4x_g\beta_gm^2((1-x_g)^2-\xi^2(1-2x_g))-\alpha_g(1-x_g^2-\xi^2)(1-\xi^2))k_\perp^2\nn\\
&+x_g^2\alpha_g(1-x_g^2-\xi^2)\Delta^2_\perp+4x_g(4\beta_gm^2x_g\xi(1-x_g)-\alpha_g(\xi+(x_g^2-\xi^2+1)))k_\perp\cdot\Delta_\perp\nn\\ &-4m^2\alpha_g((1-x_g)^2-\xi^2)^2\Big]\\
G_{14}^g=&-\frac{\alpha_g\sqrt{1-\xi^2}}{2x_g}\Big[4(1-x_g^2-\xi^2)(1-\xi^2)k_\perp^2-x_g^2(1-x_g^2-\xi^2)\Delta^2_\perp+4x_g\xi(1-x_g^2-\xi^2)\nn\\
&k_\perp\cdot\Delta_\perp+4m^2((1-x_g)^2-\xi^2)^2\Big] \label{eq: gluon G14}
  \end{align}
\end{itemize}

\bibliography{RefGTMD.bib}
\bibliographystyle{hieeetr}
\end{document}